\documentclass[traditabstract,longauth]{aa} 
\usepackage{graphicx}
\usepackage{txfonts}
\usepackage{longtable}
\begin{document}
   \title{When Nature Tries to Trick Us} \subtitle{An eclipsing eccentric close binary superposed on the central star of the planetary nebula M3-2\thanks{Based on ESO observations made under programmes 088.D-0573(A), 090.D-0435(A), 090.D-0693(A), 091.D-0475(A), 092.D-0449(A), 094.D-0031(A), 094.D-0031(A), and 096.D-0237(A).}
   }

   \author{Henri M. J. Boffin
   \inst{1}
          \and
   David Jones
   \inst{2,3}
    \and
      Roger Wesson
      \inst{4}
\and
    Yuri Beletsky
      \inst{5}
          \and
    Brent Miszalski
  	  \inst{6,7}
	  \and \\
    Ivo Saviane
   \inst{8}
   \and
   Lorenzo Monaco
    \inst{9}
    \and
    Romano Corradi
    \inst{10,2} 
    \and
    Miguel Santander Garc\'\i a
    \inst{11}
    \and
    Pablo Rodr\'\i guez-Gil
    \inst{2,3}   
          }

   \institute{European Southern Observatory, Karl-Schwarzschild-str. 2, 85748 Garching, Germany\\
\email{hboffin@eso.org}
\and
Instituto de Astrofísica de Canarias, Vía Láctea s/n, E38200, La Laguna, Tenerife, Spain
\and 
Departamento de Astrof\'isica, Universidad de La Laguna, E-38206 La Laguna, Tenerife, Spain
\and
Department of Physics and Astronomy, University College London, Gower St, London WC1E 6BT, UK
\and
Las Campanas Observatory Carnegie Institution of Washington, La Serena, Chile
\and
South African Astronomical Observatory, PO Box 9, Observatory, 7935, South Africa
\and
Southern African Large Telescope Foundation, PO Box 9, Observatory, 7935, South Africa           
\and
European Southern Observatory, Alonso de Cordova 3107, Casilla 19001, Santiago, Chile
\and
Departamento de Ciencias Fisicas, Universidad Andres Bello, Fernandez Concha 700, Las Condes, Santiago
\and
GRANTECAN, Cuesta de San José s/n, E-38712, Bre\~{n}a Baja, La Palma, Spain
\and
Observatorio Astron\'{o}mico Nacional (OAN-IGN), C/Alfonso XII, 3, E-28014 Madrid, Spain
}

   \date{Received ; accepted }

 
  \abstract
   {Bipolar planetary nebulae (PNe) are thought to result from binary star interactions and, indeed, tens of binary central stars of PNe have been found, in particular using photometric time-series that allow detecting post-common envelope systems. Using photometry at the NTT in La Silla we have studied the bright object close to the centre of PN M3-2 and found it to be an eclipsing binary with an orbital period of 1.88 days. However, the components of the binary appear to be two A or F stars, of almost equal masses, and are thus too cold to be the source of ionisation of the nebula. Using deep images of the central star obtained in good seeing, we confirm a previous result that the central star is  more likely a much fainter star, located 2\arcsec\ away from the bright star. The eclipsing binary is thus a chance alignment on top of the planetary nebula. We also studied the nebular abundance and confirm it to be a Type I PN.   
   }


\authorrunning{H.M.J. Boffin et al.}
\titlerunning{An eclipsing binary superposed on M3-2}
   \maketitle
%
   
\section{The bipolar nebula M3-2}
Planetary nebulae (PNe) are thought to be short episodes at the end of the lives of intermediate-mass stars, just before they become white dwarfs.  
Most planetary nebulae come, however, in many different shapes, which is difficult to explain in the case of the evolution of a single star \citep{2002ARAA..40..439B}. This is particularly the case for bipolar planetary nebulae, for which there is now mounting evidence that they originate from a binary system \citep{Miszalski2009, 2009PASP..121..316D, 2015ASPC..493..527B, 2015ebss.book..153B, 2017NatAs...1E.117J, MyCn18}. And indeed, many close binaries are now found at the centre of planetary nebulae\footnote{See the latest list at \url{http://drdjones.net/?q=bCSPN}.}.


\object{PN M3-2} (PN G240.3-07.6, ESO 428-5;  \citealt{{Minkowski1948}}) is a bipolar Type I planetary nebula, with a well defined ring (Fig.~\ref{fig:images}), and as such is a perfect contender to host a binary system, according to the observed correlation between nebular morphology and central star binarity \citep[][ and references therein]{Miszalski2009,jones15}. Its distance is poorly known: 
\citet{1971ApJS...22..319C} report a distance of 4.96 kpc,  \citet{1984AAS...55..253M} quotes a value of 3.2 kpc, while later values vary between 4.65 kpc and 12.42 kpc \citep{2004MNRAS.353..589P}.  \citet{2012AstL...38..707K} hypothesises that M3-2 is a possible PN belonging to the dwarf galaxy remnant in Canis Majoris, at a distance of about 7.2 kpc. The star apparently ($\alpha=$~07:14:49.92; $\delta=-$27:50:23.21 --  J2000) at the centre of the planetary nebula, although slightly offset, is rather bright, $B$=16.88, $V$=16.96  \citep{1989ApJS...69..495S}, or  $B = 16.53, V=16.31$ \citep{1991AAS...89...77}, and not as blue as a typical PN central star.

The bipolar nature of the nebula  led us to include M3-2 on our list of planetary nebulae to be followed-up for binarity. We therefore did time-resolved photometry of the object (Sec.\ref{sec:binary}), and once the binarity of the apparent central star was confirmed, we took spectra of the object (Sec.\ref{sec:spec}) and analysed the nebular abundance (Sec.\ref{sec:neb}). The outcome of our analysis is further discussed in Sec.\ref{sec:dis}.

\section{Imaging and stellar photometry}\label{sec:binary}
\subsection{An intriguing binary}
M3-2 was initially observed during a five-day campaign using ESO's 3.58-m New Technology Telescope equipped with the EFOSC2 instrument \citep{efosc2}, between 27 February and 3 March 2012. Observations were done in the Gunn $I$-band, using the i\#705 filter ($\lambda_0 = 793.1$ nm, $\Delta \lambda = 125.6$ nm), with an exposure time of 30 seconds.  Exposure overheads, including CCD read-out time, limited the time resolution to no better than 65 s. Initially, observations were done in blocks of 10 photometric points (i.e. for about 10 minutes), repeated two to four times per night (with the blocks distributed relatively evenly throughout the observing window for the object) during the three first nights. The data were pre-reduced and the light curve estimated on the spot, to make sure that any variability could be detected in real time. Although the flux remained constant during the almost four hours spanned by the observations done during the first night, as well as for our first epoch on the second night, the variable nature of the object was suddenly revealed during the second epoch of the same night as the flux dropped by more than $0.4$~mag. On the third night, the flux was back to the initial levels, despite returning to the object four times during the night, for a total time of a little over four hours. Luckily, our first epoch of the fourth night showed that the flux had dropped again, and we therefore stayed on target for 1$^{\rm h}$41$^{\rm m}$, clearly seeing that we had reached the bottom of what was obviously an eclipse, and that the flux rose again. We then made four more excursions to the object afterwards, trying to sample as uniformly as possible the eclipse egress. As this programme was aimed at discovering many new binary central stars of planetary nebulae, it would not have been efficient to remain on target for the whole time. The following night, a new eclipse could be followed, and we did regular, short excursions to the object, so as to cover different parts of what we thought was a single eclipse, which was already well sampled.
In total, on this first run, we obtained 253 data points over the course of five nights.  

Over the next four years, we returned regularly to the target, sampling the entirety of the light curve and obtaining in total 603 data points, for a total time on target of almost 11 hours. The
resulting photometric measurements are available in the online
data  (see Tab.~\ref{table:obs}). For the final data reduction, all frames have been bias-subtracted and flat-fielded using {\tt STARLINK} routines \citep{STARLINK}. Flat fields were obtained regularly on sky during twilight. Differential photometry was performed using more than ten field stars as comparison, using the same methodology as used by \citet{jones14}, i.e. using {\tt sextractor} and an aperture radius of 3$\arcsec$ ($\sim$1.5 times the worse seeing).

In addition, we have also obtained images of the PN M3-2 in different broad- and narrow-band filters:
$B$ \#639 ($\lambda_0 = 440.0$ nm , $\Delta \lambda = 94.5$ nm), $H_\alpha$ \#692 ($\lambda_0 = 657.7$ nm , $\Delta \lambda = 6.2$ nm), $H_\beta C$ \#743, [$\ion{O}{iii}$] \#687  ($\lambda_0 = 500.4$ nm , $\Delta \lambda = 5.6$ nm), and [$\ion{S}{ii}$] \#700 ($\lambda_0 = 673.0 $ nm , $\Delta \lambda = 6.2$ nm).  Images based on all these filters and showing the whole EFOSC2 4.12$\times$4.12\arcmin field of view are shown in  Online Fig.~\ref{fig:images}.

   \begin{figure}
   \centering
   \includegraphics[width=9cm]{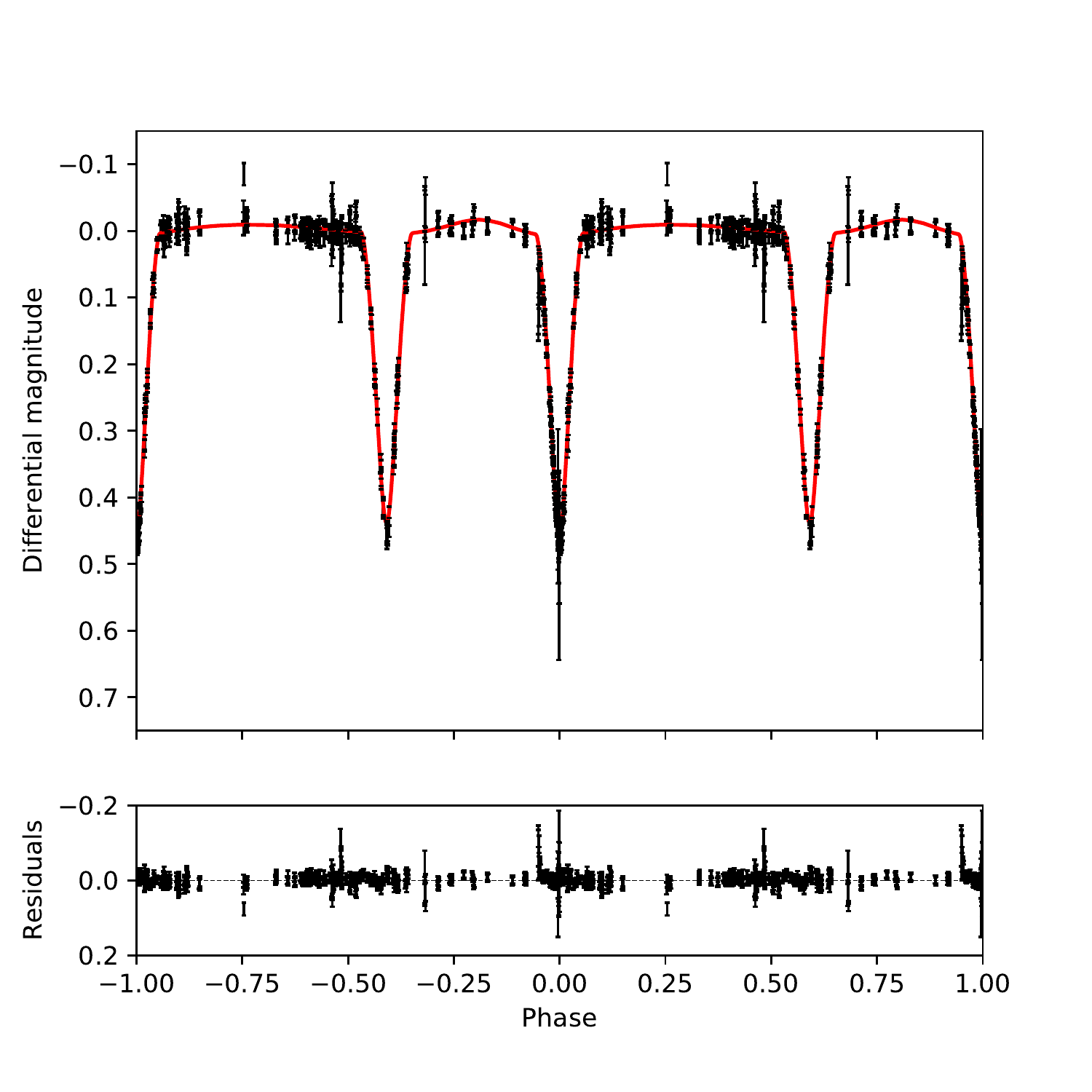}
      \caption{Phase-folded light curve of M3-2, assuming a period of 1.8767 days. The fit based on the parameters presented in Table~\ref{table:fit} is indicated by the red curve. The differential magnitude was arbitrarily normalised.}
         \label{fig:folded}
   \end{figure}
   
\subsection{An eclipsing binary... composed of two A stars}
Our light curve clearly revealed the presence of deep eclipses. During our first five-night run, we covered three times the (almost) bottom of an eclipse. 
However, while the last two were separated by 1.11 days, this would not fit with the difference between the two first ones. Our data thus indicate that we are not witnessing one but two eclipses per period, which has to be close to 1.874 days. This provided a reasonable phase-folded light curve, with two almost equal eclipses. 


\begin{table}
\caption{Best fit model of the binary system}             
\label{table:fit}      
\centering                          
\begin{tabular}{ll}        
\hline\hline                 
Parameter &  Value\\    
\hline                        
Orbital period (days)      &   1.8767113 $\pm$   0.0000013\\
T$_0$  (HJD) &      2455988.614835$\pm$   0.000215\\
Orbital inclination     &    84.2$^{+0.6}_{-0.3}$\\
Eccentricity & 0.149$\pm$0.002\\  
Omega  (degrees):       &                   16.6$\pm$3.0 \\
Phase of primary eclipse&                0.000 \\
Phase of first quadrature&                 0.295\\
Phase of secondary eclipse&                0.591\\
Phase of second quadrature&                0.795\\
Phase of periastron&                     0.84\\
Phase of apastron&                         0.34\\
Primary temperature (K)	&					8120$^{+290}_{-180}$\\
Primary radius (R$_\odot$)&                 1.7$^{+0.1}_{-0.2}$\\
Primary mass (R$_\odot$ &					1.2$\pm$0.1\\
Secondary temperature (K) &					8160$^{+280}_{-160}$\\
Secondary radius (R$_\odot$)&               1.5$^{+0.2}_{-0.1}$\\
Secondary mass (R$_\odot$)&					1.3$\pm$0.1\\
\hline                                   
\end{tabular}
\end{table}

The resulting phase-folded light curve of all our data, based on a periodogram analysis, is shown in Fig.~\ref{fig:folded}. The eclipses are clearly not separated by 0.5 orbital cycle, but by about 0.6. This implies that the binary orbit is eccentric.

   \begin{figure*}
   \centering
   \includegraphics[width=15cm]{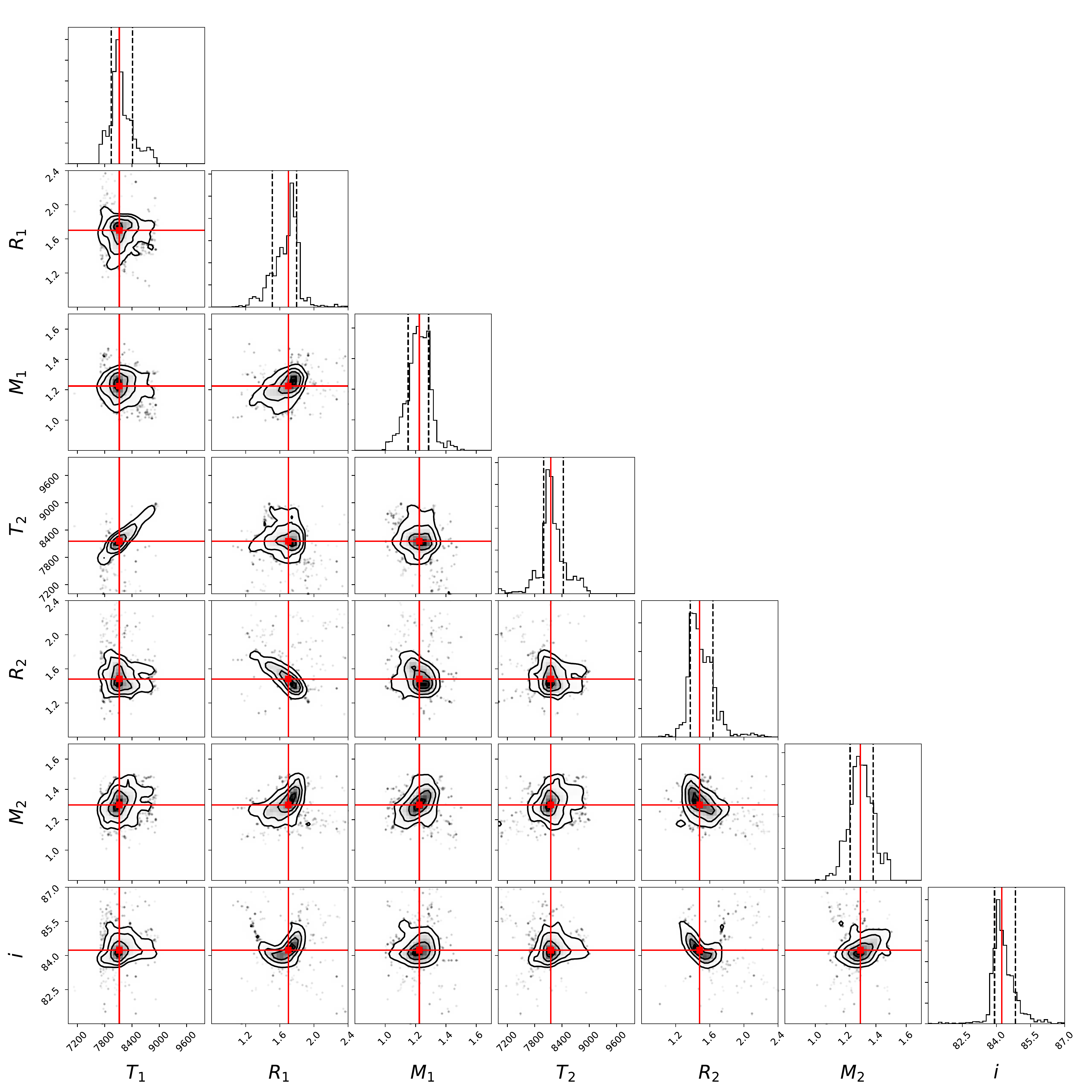}
      \caption{A corner plot for the {\tt PHOEBE 2.0} MCMC fit to the light curve. Indicated are the temperature, radius and mass of the two stars, as well as the orbital inclination. The red lines represent the most likely values, while the dashed lines reflect the  one sigma limits.}
         \label{fig:corner}
   \end{figure*}
   
We model the light curve using the  {\tt PHOEBE 2.0} code \citep{2016ApJS..227...29P} using a Markov chain Monte Carlo (MCMC) method implemented via {\tt emcee} \citep{2013PASP..125..306F} and parallelised to run on the LaPalma supercomputer with {\tt schwimmbad} \citep{2017JOSS....2..357P}.  We allowed the masses, temperatures and radii of both stars as well as the orbital inclination to vary.  The limb-darkening values for both stars were fixed to the default prescription in {\tt PHOEBE 2.0}, whereby values for each point on the star are interpolated from tables derived from stellar atmosphere model emergent intensities calculated for 32 points along the stellar limb as described in \citet{2016ApJS..227...29P}.  A corner plot of the resulting MCMC chain for the varied parameters is shown in Fig.~\ref{fig:corner} while the parameters of the resulting fit are given in Tab.~\ref{table:fit}.

As can be seen from Fig.~\ref{fig:folded}, the fit is very good, with the r.m.s. being of the order 15 mmag. The eccentric orbit is confirmed, with a value of $e$=0.149. Thus, even if the orbital period is well within the range of observed and expected periods for close binary central stars of PNe, such a finite eccentricity is rather puzzling for the outcome of a common-envelope evolution, which should in most cases (if not always) lead to a circular orbit.

Given the lack of colour information provided by the single-band light-curve, the individual temperatures of the two stars are relatively poorly constrained -- with a strong correlation between the temperatures of the primary and the secondary.  However, as implied directly from the similar eclipse depths, the stars are found to present extremely similar temperatures of around 8,100--8,200 K (corresponding to an A-type star).  The radii and masses are slightly better constrained but, once again, show clear correlations between the two stars with the mass and radii ratios both being roughly unity.  The inclination of the orbit is better constrained at  $i$=84.2$^\circ$.

\begin{figure*}
   \centering
   \includegraphics[width=18cm]{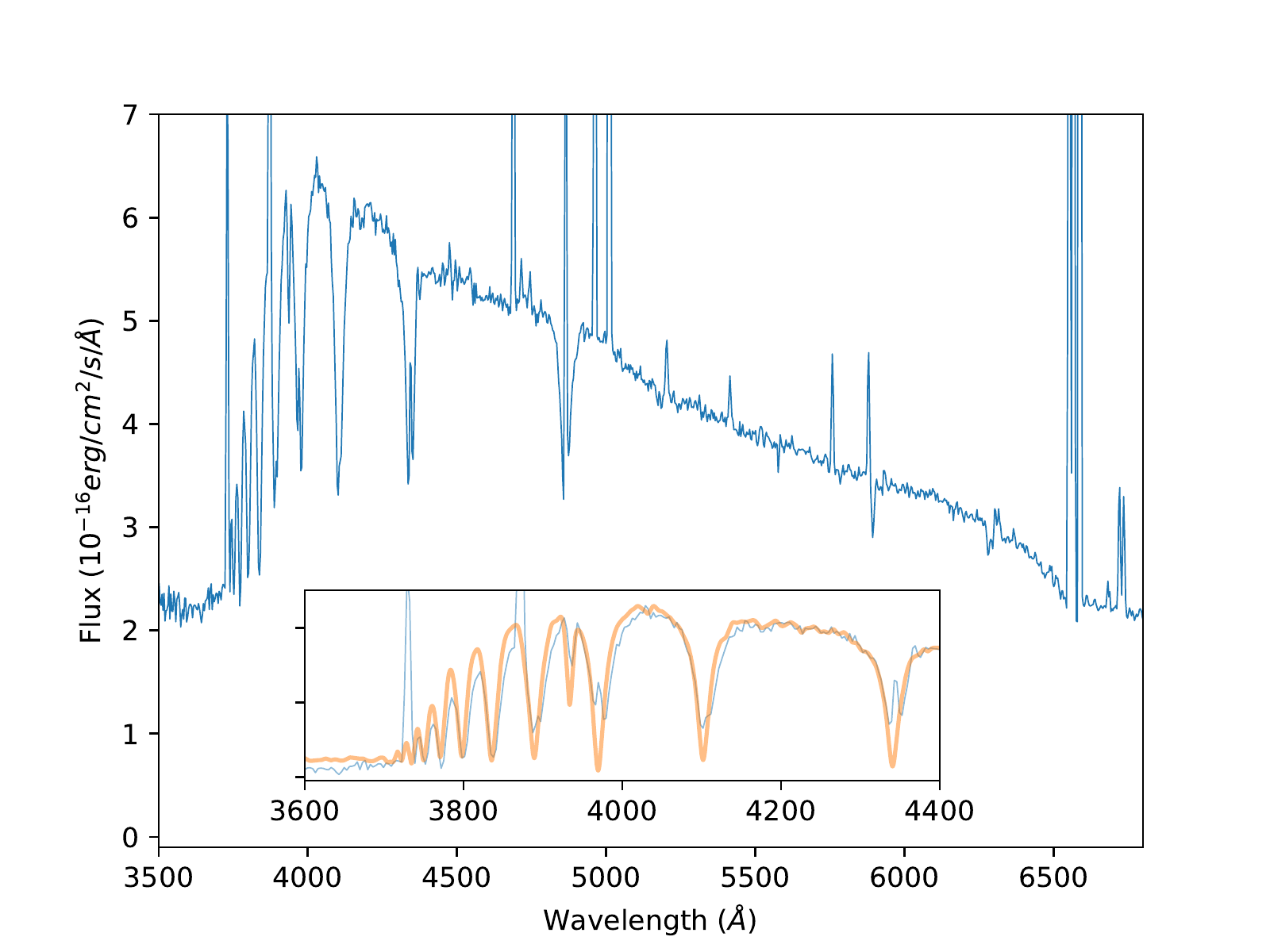}
      \caption{FORS2 spectrum of the bright central star of M3-2 taken with the 300V grism. No attempt was made to remove the nebular lines. The inset shows a zoom in the blue part of the spectrum, with in  orange  a synthetic spectrum corresponding to a model of T$_{\rm eff}$=8,500 K, $\log g=4.0$. As the stellar spectrum is affected by the nebular lines, the bottom of the lines are not well fitted. }
         \label{fig:spec}
   \end{figure*}

\section{Stellar spectroscopy}\label{sec:spec}   
In order to better characterise the newly-discovered binary system, we obtained on the night of 7--8 January 2016 a spectrum of the central star, using FORS2 and the 300V grism, with a dispersion of 112 \AA/mm and covering the wavelength range 3300 -- 6600 \AA, above which second order contamination is present. The slit  was aligned North-South, had a width of 0.7\arcsec\ and the exposure time was 180s.  The seeing was about 0.8\arcsec, while the observations were done at airmass 2. The white dwarf GD50 was used as a spectrophotometric standard. The resulting spectrum is shown in Fig.~\ref{fig:spec}.
The obtained spectrum corresponds to an orbital phase of $\phi = 0.41$ according to our ephemeris, i.e. out of eclipse. 
 
Given the results from the light curve fitting, we have fitted the observed stellar spectrum assuming two similar objects, using the stellar spectral synthesis program {\tt SPECTRUM}\footnote{{\tt SPECTRUM } is available at \url{http://www.appstate.edu/~grayro/spectrum/spectrum.html}.} \citep{1994AJ....107..742G}. We neglected here the extinction, as it is known to be very small (see below). In order to fit correctly the Balmer jump as well as the wings of the Balmer series, we find that models with effective temperature T$_{\rm eff}$= 8,000 -- 8,500 K, and gravity $\log g=4.0 - 4.5$ are needed. This provides an independent confirmation of our light curve analysis. 

\begin{figure*}
   \centering
   \includegraphics[width=18cm]{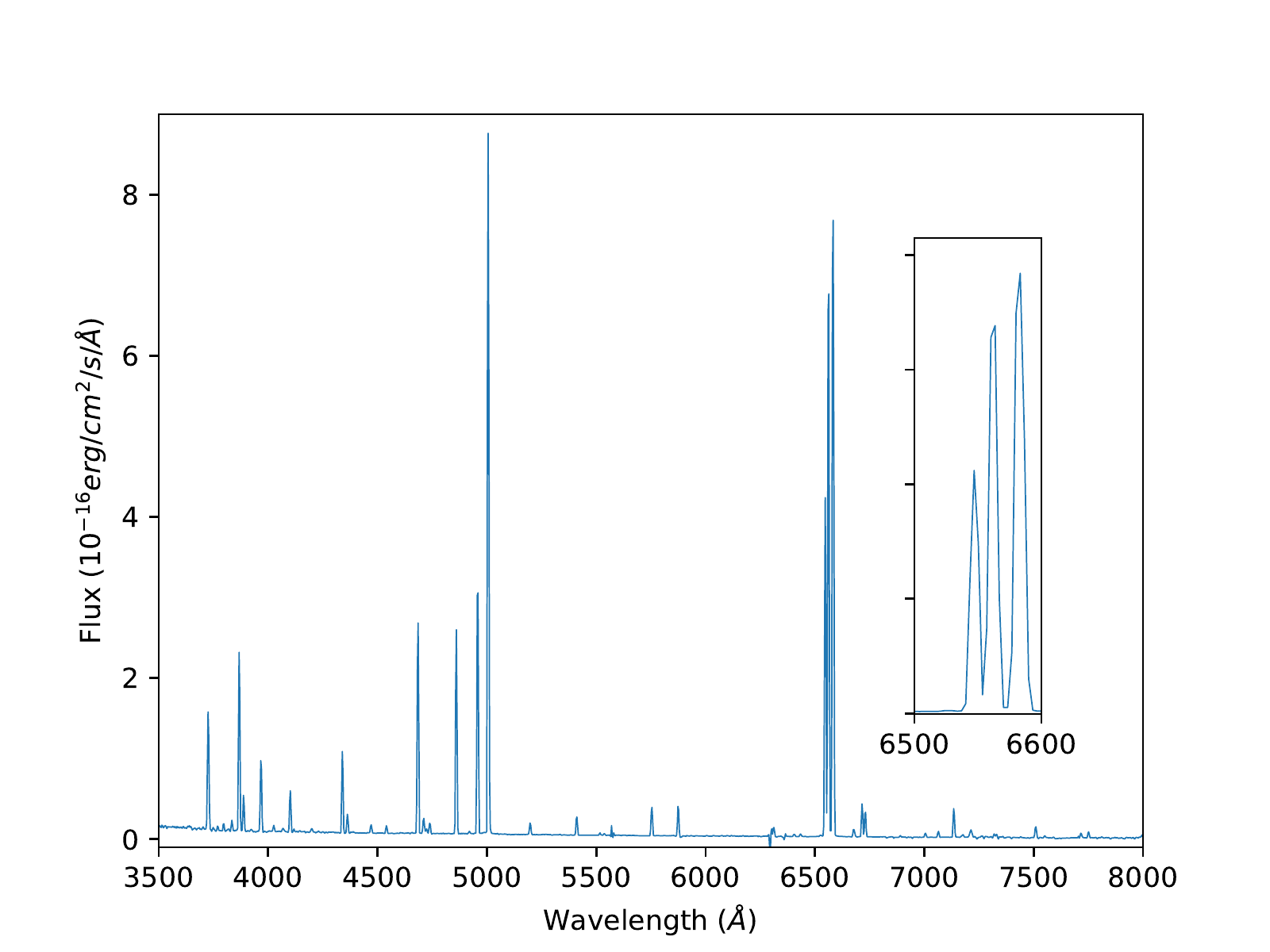}
      \caption{FORS2 spectrum of the nebula of M3-2 taken with the 300V grism, a 1\arcsec\ wide slit and an exposure time of 1800s. The right inset shows the [N\ion{II}] and H$\alpha$ lines, illustrating the strong nitrogen lines.}
         \label{fig:nebspec}
   \end{figure*}

\section{Nebular abundances}\label{sec:neb}

In addition to the stellar spectrum, we have also secured on the night of 6--7 January 2016, a deep (1800s) spectrum of the nebula of M3-2 using FORS2, the same 300V grism and same slit width of 0.7\arcsec.  The seeing was also around 0.8\arcsec, the observations were done at airmass 1.3, and the slit was also oriented North-South. The same spectrophotometric standard star was used. The spectrum is shown in Fig.~\ref{fig:nebspec}. The slit was placed such as to avoid the bright star,  3\arcsec to the East.

From the ratio of H$\alpha$ to H$\beta$, we derive an extinction c(H$\beta$) = $0.093\pm0.069$, giving $A(V)=0.19$. This value is in agreement with \cite{2012AstL...38..707K}, who found  c(H$\beta$) = $0.12\pm0.05$. 

We used {\tt alfa} \citep{Alfa} to fit the nebular lines and {\tt neat} \citep{Neat} to determine the abundances of the chemical elements\footnote{Both software are available for download from \url{https://github.com/rwesson}.}.  The effectiveness of {\tt alfa} and {\tt neat} in studying the physical and chemical properties of planetary nebulae from FORS2 spectroscopy has been well demonstrated by previous studies \citep[e.g.,][]{jones16}. {\tt Alfa} optimises Gaussian fits to the observed emission lines using a genetic algorithm. The line fluxes thus measured are then passed to {\tt neat}, which applies an empirical scheme to calculate the abundances. The calculations use a three zone model of low, medium and high ionisation. Uncertainties are propagated through all steps of the analysis into the final values. 

The resulting line intensities are shown in the Online Tab.~\ref{Tab:lines}, while the outcome of {\tt neat} is shown in Tab.~\ref{Tab:neat}. Our results confirm the values obtained by  \cite{2012AstL...38..707K}, namely that the PN M3-2 is a Type I PN with a rather low oxygen abundance but a high helium content. The logarithmic abundance of [O/H] is 7.8, considerably lower than the value of 8.54 found for a sample of PNe outside the solar circle \citep{2005MNRAS.362..424W}.


 \begin{table}
 \caption{\label{Tab:neat} Output of {\tt neat} abundance analysis of M3-2}
 \begin{tabular}{ll}
 \hline\hline
 Parameter & Value\\
 \hline
c(H$\beta)$ (H$\alpha$/H$\beta$)      & ${  0.09}^{+  0.07}_{ -0.07}$ \\
 \vspace{0.02cm}\\
{}[O~{\sc ii}] density              & ${  1970}^{+   220}_{  -200}$ \\
{}[N~{\sc ii}] temperature          & ${ 12700}\pm{   200}$ \\
{}[Cl~{\sc iii}] density            & ${  4970}^{+  6560}_{ -3470}$ \\
Medium ionisation density           & ${  4610}^{+  6480}_{ -3000}$ \\
{}[O~{\sc iii}] temperature         & ${ 16200}\pm{   700}$ \\
Medium ionisation temperature       & ${ 16200}\pm{   700}$ \\
 \vspace{0.02cm}\\
 N$^{+}$/H                           & ${  1.11\times 10^{ -4}}\pm{  4.00\times 10^{ -6}}$ \\
icf(N)                              & ${  2.99}^{+  0.23}_{ -0.19}$ \\
N$^{}$/H                            & ${  3.33\times 10^{ -4}}^{+  2.30\times 10^{ -5}}_{ -1.90\times 10^{ -5}}$ \\
O$^{+}$/H                           & ${  2.03\times 10^{ -5}}^{+  1.20\times 10^{ -6}}_{ -1.10\times 10^{ -6}}$ \\
O$^{2+}$/H                          & ${  3.56\times 10^{ -5}}^{+  3.90\times 10^{ -6}}_{ -3.10\times 10^{ -6}}$ \\
icf(O)                              & ${  1.19}^{+  0.02}_{ -0.03}$ \\
O$^{}$/H                            & ${  6.66\times 10^{ -5}}^{+  5.10\times 10^{ -6}}_{ -4.20\times 10^{ -6}}$ \\
Ne$^{2+}$/H                         & ${  2.21\times 10^{ -5}}^{+  2.80\times 10^{ -6}}_{ -2.20\times 10^{ -6}}$ \\
icf(Ne)                             & ${  1.20}^{+  0.02}_{ -0.03}$ \\
Ne$^{}$/H                           & ${  2.63\times 10^{ -5}}^{+  3.40\times 10^{ -6}}_{ -2.70\times 10^{ -6}}$ \\
Cl$^{2+}$/H                         & ${  4.42\times 10^{ -8}}^{+  1.30\times 10^{ -8}}_{ -1.06\times 10^{ -8}}$ \\
icf(Cl)                             & ${  1.58}^{+  0.03}_{ -0.04}$ \\
Cl$^{}$/H                           & ${  6.97\times 10^{ -8}}^{+  2.19\times 10^{ -8}}_{ -1.78\times 10^{ -8}}$ \\
 \vspace{0.02cm}\\
He$^{+}$/H                          & ${  0.14}^{+  0.02}_{ -0.01}$ \\
He$^{2+}$/H                         & ${  0.05}\pm{  1.00\times 10^{ -3}}$ \\
He/H                                & ${  0.18}^{+  0.02}_{ -0.01}$ \\
\hline
 \end{tabular}
 \end{table}

\begin{figure}
   \centering   \includegraphics[width=\hsize]{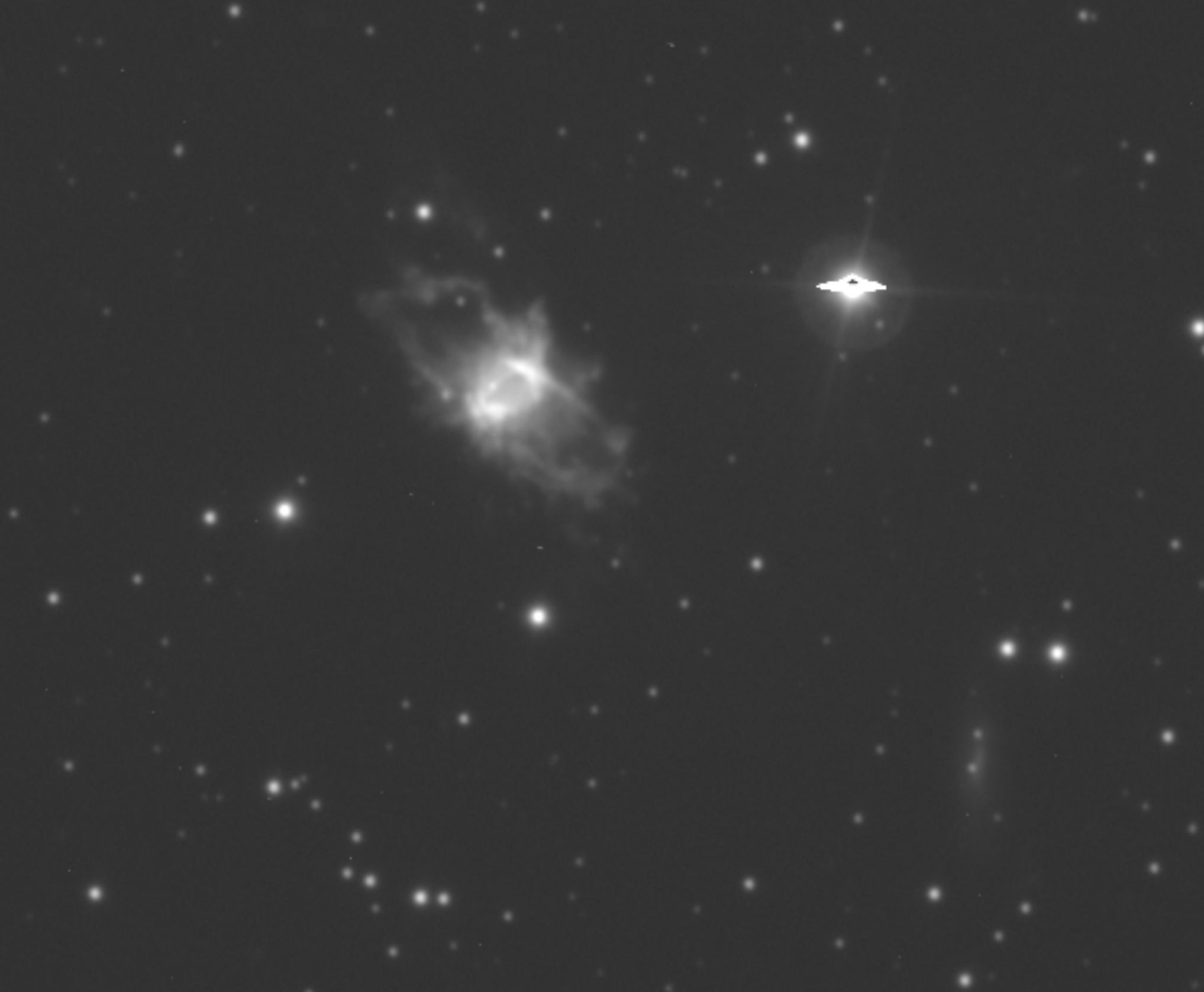}
      \caption{M3-2 in H$\alpha$ observed with FORS2. The field of view is 2.4'$\times$1.9'. North is up and East is to the right.}
         \label{fig:FORS2Ha}
   \end{figure}

  \begin{figure}
   \centering
   \includegraphics[width=\hsize]{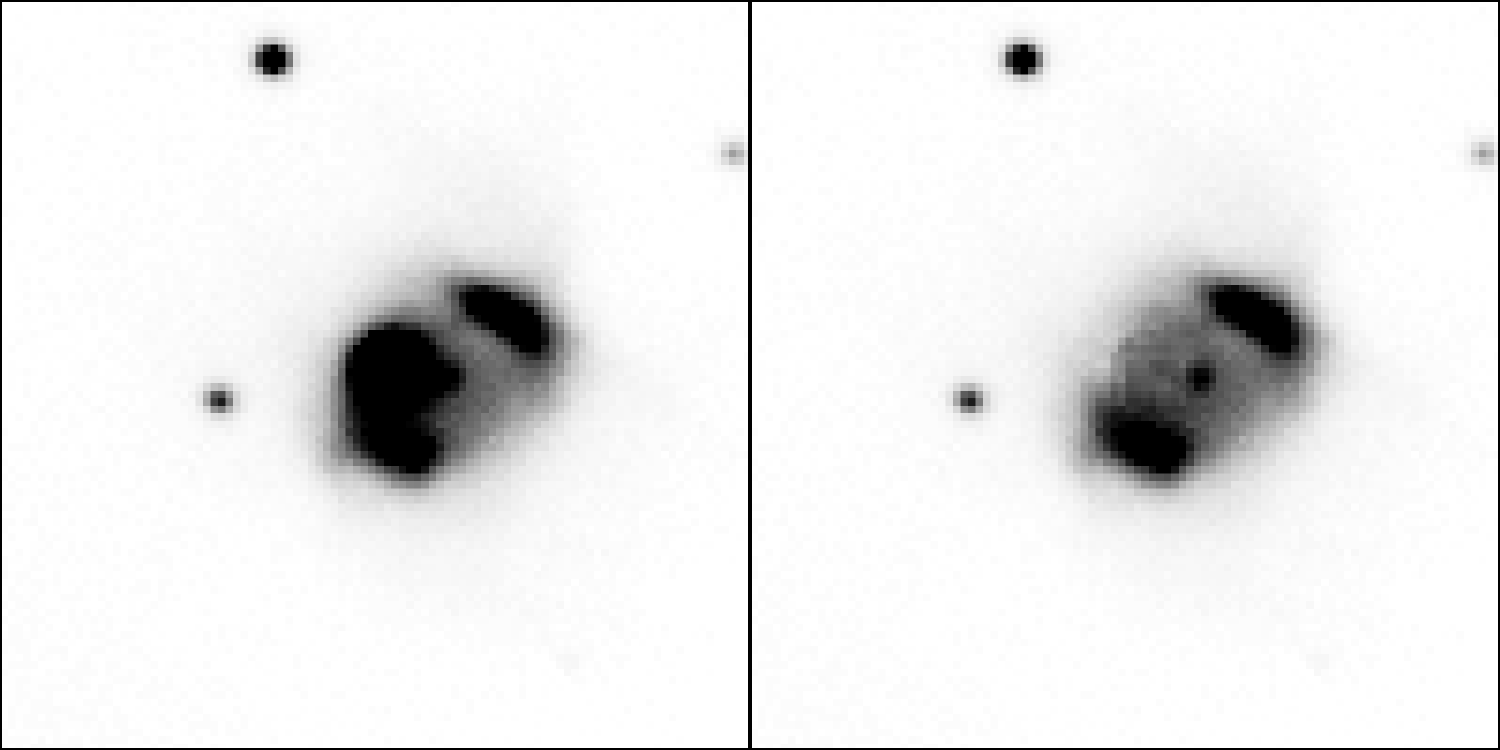}
      \caption{Showing the likely CSPN after substracting the bright object.}
         \label{fig:cspn2}
   \end{figure}

\section{Analysis: A clear case of false identity}\label{sec:dis}
The object that we observed, despite its projected position appearing close to the centre of the nebula, is most likely not the real central star of M3-2. Indeed, as it consists of a close binary with an eccentric orbit containing two main-sequence stars, of A or F spectral types, it cannot be the progenitor of the ionised nebula. 

The fact that the bright object is not the real CSPN should not come as a surprise, though. First, it is obviously off-centre, although there are some other examples of PN with off-centre CSPN. Second, based on the Ambartsumyan (or ``crossover'') temperature of the central star of 267,000 K, and given its distance, \cite{1989ApJ...345..871} found that the hot central star of M3-2 should have a $V$-magnitude of  23.36, well below the observed value, leading these authors to state that ``{\it clearly we have measured the bright companion of the true central star}''. The same authors \citep{1989AJ.....98.1662} seem to have found a much fainter star ($V=21.1$) at the centre of the PN, $\sim$3\arcsec\ from the bright central star. 
They measured its magnitude  after removing the bright star by subtracting a scaled stellar profile from the image of the star.  

To confirm this, we obtained deep images of the PN M3-2 with FORS2 in both H$\alpha$ and $B$-band. They are shown in Fig.~\ref{fig:FORS2Ha} and  Fig.~\ref{fig:cspn1}, respectively. While  Fig.~\ref{fig:FORS2Ha}  clearly reveals the intricate nature of the nebula and its bipolar shape,  Fig.~\ref{fig:cspn1} allows us to detect a faint object next to the bright star we have been analysing till now. By doing a point spread function (PSF) fitting of the bright star and subtracting it from the image, a faint star at the very centre of the nebula is clearly visible (see Fig.~\ref{fig:cspn2}) and is most likely the real CSPN, especially as it appears very blue (although its exact colours are rather uncertain due to the nebular contamination and possible remnant artefacts from the subtraction of the bright binary). The estimated brightness and location are provided in Tab.~\ref{Tab:real}. 

\begin{table}
\caption{\label{Tab:real} Properties of the  likely CSPN of PN M3-2}
\begin{tabular}{ll}
\hline \hline
Parameter & Value\\
 \hline
Right Ascension (J2000):& 07:14:49.77\\
Declination (J2000):&-27:50:23.59\\
$B$ mag. on MJD 57394.3290944&    $\approx 20.3$\\     
$V$ mag. on  MJD 57393.1836432&    $\approx 21.7$   \\ 
\hline
\end{tabular}
\end{table}

\cite{1991AAS...89...77} measured the bright object to have magnitudes $B = 16.53$ and $V=16.31$, presenting a colour excess $B-V$ consistent with the range of temperatures derived by our light-curve modelling \citep{1998PASP..110..863P}. Taking an absolute magnitude and bolometric $V$-band correction from \citet{1998PASP..110..863P}, the above-derived $A(V)=0.2$, and accounting for the fact that the observed apparent magnitude corresponds to the sum of both stars, implies a distance to the binary of approximately 8 kpc. Similarly, based on the comparison between our synthetic spectra and the flux calibrated FORS2 stellar spectrum, we are led to a ratio between the distance and the stellar radius of $(d/R) = (2.75 \pm 0.22)~ 10^{11}$. Given the radius determined from the light curve, this translates to a distance of $7.5 \pm 0.6$ kpc, depending on the effective temperature and gravity. This distance is well beyond the estimated one to the nebula, in most cases, between 3 and 5 kpc (see above), although it is compatible with some values provided in the literature. The binary system and the nebula are thus most likely not linked, except by some unfortunate and distracting alignment. 
Given that the binary is clearly not containing the CSPN, we did not follow it sufficiently to derive the spectroscopic orbit and thus determine its systemic radial velocity. If done, this could be compared against the nebula's velocity as an additional check. Based on our analysis of the nebular lines, the radial velocity of the PN is $66.7\pm3.9$~kms$^{-1}$.
If the binary and the nebula were linked, however, as the separation on sky is 2\arcsec, the physical separation would be $\approx15,000$ au, making it a very wide, most likely unbound, hierarchical triple.
We note that the Gaia DR2 release\footnote{\url{https://gea.esac.esa.int/archive/}} \citep{gaia16,gaia18} lists a star of $G=16.239$ and colour $G_p-R_p=0.48$ at the position of the bright star ($\alpha=$~07:14:49.92; $\delta=-$27:50:23.2;  J2000). The parallax is, however, not very useful for now, being 0.0535$\pm$0.0516 mas, but would be in line with a very distant object, and should, hopefully, be more precise in future releases. Given its brightness, the real CSPN, though, will not be available in any future Gaia releases.

\section{Discussion and conclusions}
A photometric monitoring of the PN M3-2 reveals the presence of an eclipsing close binary with an orbital period of 1.88 days in an eccentric orbit. Because close binary CSPNe are the result of common-envelope evolution -- a strongly dissipative process -- mostly circular orbits are expected, and this result is therefore at first sight very surprising. Furthermore, the light curve shows the presence of two almost equal eclipses, whose depth and duration imply the presence of stars with an almost solar radius size. Additional spectroscopy confirms that the components of the system are almost equal-mass main-sequence stars, located about 8~kpc away. This is much farther than the estimated distance of the planetary nebula, showing that this binary is most likely just a chance superposition onto the true CSPN. The CSPN itself is  much fainter, most likely located 2\arcsec\ from the bright star and with an estimated $B \sim 20.3$. 

The mere existence of this alignment acutely illustrates the non-significance of using a statistical argument when disproving a chance alignment. The recent confirmation that the binary thought to be related to the CSPN of \object{PN SuWt 2} is merely a field star system, by chance lying in the same line of sight as the nebular centre, and that it bears no relation to SuWt 2 or its, as yet unidentified, central star(s) \citep{2017MNRAS.466.2034J}, shows that this unfortunate circumstance exists now twice.
We can try to crudely quantify this by querying SIMBAD for all stars listed as having spectral type A.  This returns $\approx 100,000$ of them in total.  Then,  we assume a population of PNe of $\approx 3,000$. If two populations of objects of these sizes were uniformly distributed across the sky -- a very crude approximation -- then the probability that there is one chance alignment to within 10\arcsec\ among the 3$\times$10$^8$ possible pairs of objects is 1.4\%. The probability of two such alignments is just 0.02\%. Of course, the fact that both A stars and PNe are strongly concentrated in the Galactic plane increases the chances of such alignments, but the occurrence of two is still remarkably unlikely.

Another estimate can be done using the Gaia DR2 catalogue. We have probed for all objects within 1 degree of the PN M3-2 that are bright  ($G < 19$) and blue ($-0.1 < B_p-R_p < 0.5$). We obtain 2476 possible sources. Doing the same for PN SuWt 2 returns 420 objects. Thus, scaling the 1 degree field to a 10\arcsec\ region, leads to a 1.9\% probability of a chance alignment around M3-2 and 0.3\% around SuWt 2. Assuming both are independent, the chance to have two alignments would be 0.006\% and thus given the 3,000 PNe known, less than 0.2 occurrences should be possible. 
We do know one such occurrence, however. It thus appears clear that Nature likes to play tricks with us!

It is also worth mentioning that we also know of a few cases of superpositions of bright stars with bulge PNe (see, e.g., Appendix E of \citealt{2013MNRAS.432.3186M}), many of which will have a similar distance of $\sim8$ kpc.
 The anonymous referee asks us to mention that these results raise the possibility that other binary central stars may, in fact, be misidentified field stars -- particularly those in crowded fields (such as in the galactic bulge). However, this is mostly true for stars that haven't been studied enough, as in most cases, the colours of the central stars should already provide confirmation of their nature. Moreover, whenever possible, radial velocities, UV/NIR photometry, spectroscopic peculiarities, etc., can be used to establish an association \citep[e.g.,][]{A70,Hen2-39}.

Given the clear bipolar nature of PN M3-2, it seems likely that it contains a binary central star, and we encourage readers to try to prove this. Given the magnitude of the central star, this may not be easy though. One way forward would be to use the link between close binarity and the abundance discrepancy factor (Wesson et al., subm.), who have found that central star binary periods are correlated with other observable parameters: considering the discrepancy between recombination line and collisionally excited line abundances, extreme values ($>$10) of this discrepancy are seen only where the binary period is less than about 1.15 days. These short periods and extreme abundance discrepancies are also associated with low density ($<$1000cm$^{-3}$) nebulae. In the case of M3-2, our spectra are not deep enough to detect recombination lines of O$^{2+}$, but we can place an upper limit to the abundance discrepancy factor (\textit{adf}) of about 15. In addition, the density of the nebula is measured at $\sim$2000\,cm$^{-3}$ from [O~{\sc ii}] lines and $\sim$5000\,cm$^{-3}$ from [Cl~{\sc iii}] lines. While the upper limit to the \textit{adf} is quite weak, the density of M3-2 is higher than any of the nebulae with short period binary central stars and a measured chemistry. Thus, the morphology of the nebula points to a short period binary central star, while its chemistry and density further suggest that this binary should have a period longer than about 1.15 days.
  
\begin{acknowledgements}
Part of the work was done while HMJB was visiting the IAC, thanks to a visitor grant in the framework of a Severo Ochoa excellence programme (SEV-2015-0548). This research made use of Astropy, a community-developed core Python package for Astronomy \citep{2013A&A...558A..33A,2018arXiv180102634T}, numpy \citep{numpy}, matplotlib \citep{matplotlib} and corner \citep{2016JOSS....1...24F}. This  research  has  been  supported  by  the  Spanish  Ministry  of  Economy  and  Competitiveness  (MINECO)  under the grant AYA2017-83383-P.  The  authors  thankfully acknowledge the technical expertise and assistance provided by the Spanish Supercomputing Network (Red Espa\~nola
de Supercomputaci\'on), as well as the computer resources used:  the LaPalma Supercomputer, located at the Instituto
de Astrof\'\i sica de Canarias. B.M. acknowledges support from the National Research Foundation (NRF) of South Africa. This work was partially funded by the Spanish MINECO through project AYA2016-78994-P. RW was supported by European Research Grant SNDUST 694520. This work has made use of data from the European Space Agency (ESA) mission {\it Gaia} (\url{https://www.cosmos.esa.int/gaia}),
processed by the {\it Gaia} Data Processing and Analysis Consortium (DPAC,
\url{https://www.cosmos.esa.int/web/gaia/dpac/consortium}). Funding for the DPAC
has been provided by national institutions, in particular the institutions
participating in the {\it Gaia} Multilateral Agreement.
\end{acknowledgements}

\begin{appendix}
\section{Additional tables and figures}

 \begin{figure*}
   \centering
   \includegraphics{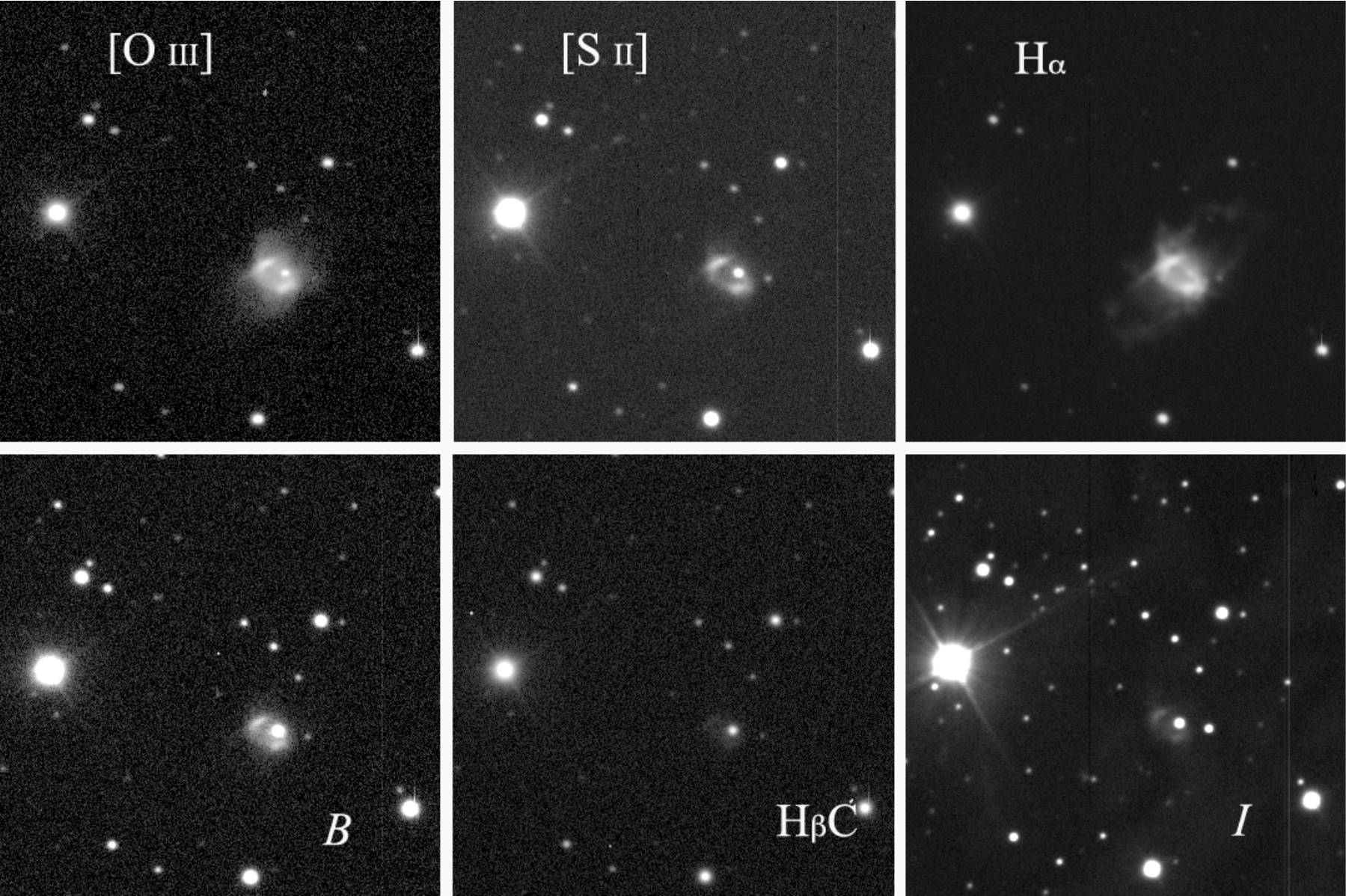}
      \caption{EFOSC2 images of M3-2 in different narrow-band (top) and broadband (bottom) filters. North is up and East is to the left.}
         \label{fig:images}
   \end{figure*}

\begin{figure*}
   \centering
   \includegraphics[width=\hsize, angle=0]{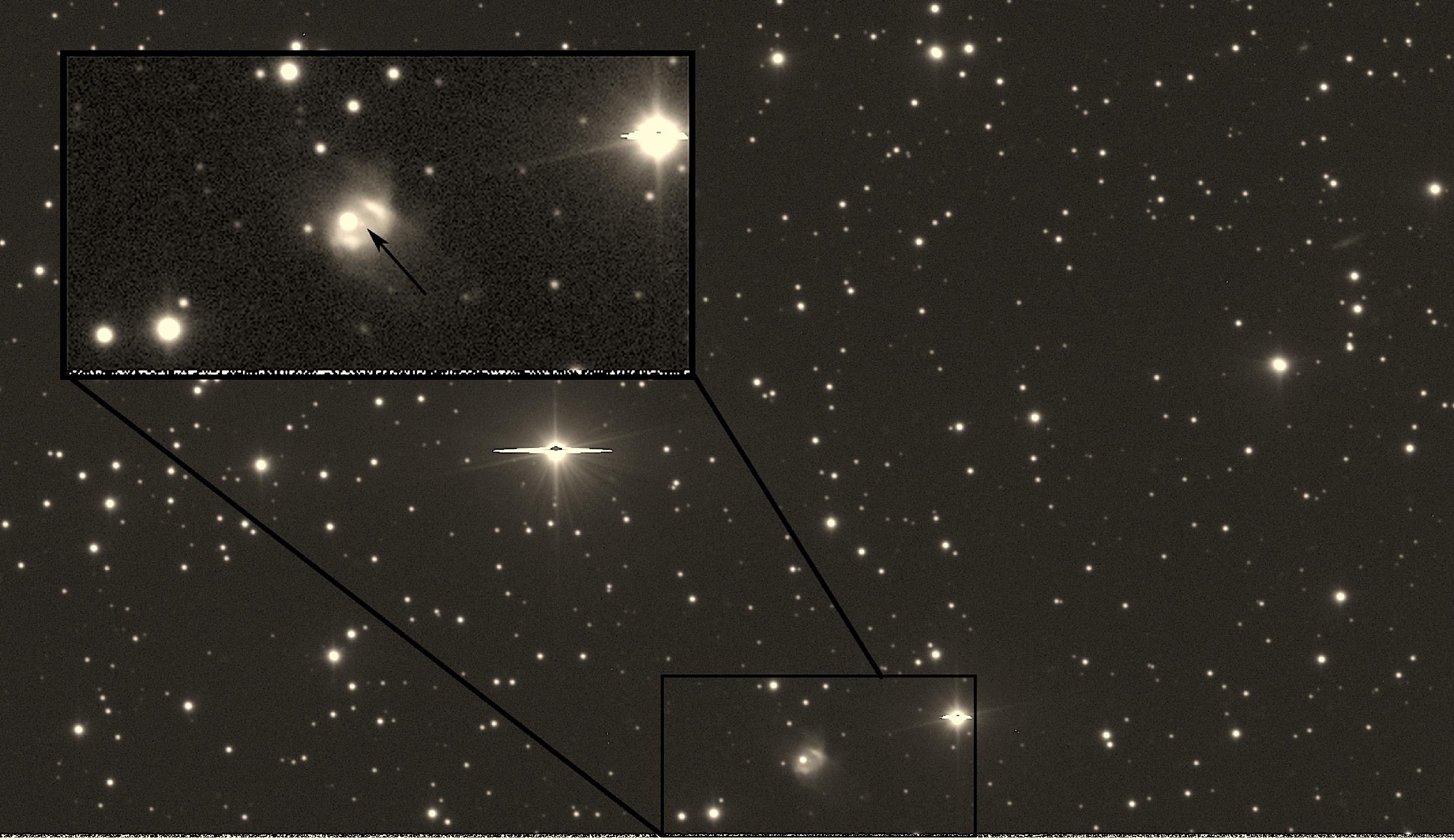}
      \caption{Showing the likely CSPN with FORS2, with a deep image in $B$.The intensity is shown with a logarithmic scale and the upper inset shows a zoom-in around the PN. The arrow indicates the real CSPN.}
         \label{fig:cspn1}
   \end{figure*}
   
\clearpage
\onecolumn
\longtab{1}{
\begin{longtable}{l c c c c} 
\caption{\label{table:obs} EFOSC photometry of M3-2. The second column is the difference between the magnitude of the bright object in M3-2 and a comparison star, while the fourth column is the difference between the magnitude of two comparison stars to serve as a check of the stability.}\\             
\hline\hline                 
Date & $\Delta$Mag. M3-2 & Error & $\Delta$Mag comp. & Error \\    
\hline                        
\endfirsthead
\caption{continued.}\\
\hline\hline
Date & $\Delta$Mag. M3-2 & Error & $\Delta$Mag comp. & Error \\  
\hline
\endhead
\hline
\endfoot
2455985.5875 & 0.5719 & 0.0096 & -0.8398 & 0.0107 \\
2455985.59087 & 0.5687 & 0.0094 & -0.8393 & 0.0106 \\
2455985.59162 & 0.5665 & 0.0095 & -0.8323 & 0.0107 \\
2455985.59237 & 0.5726 & 0.0095 & -0.8330 & 0.0106 \\
2455985.59312 & 0.5701 & 0.0095 & -0.8330 & 0.0106 \\
2455985.59388 & 0.5762 & 0.0095 & -0.8407 & 0.0106 \\
2455985.59463 & 0.5749 & 0.0095 & -0.8400 & 0.0106 \\
2455985.59537 & 0.5726 & 0.0095 & -0.8377 & 0.0107 \\
2455985.59611 & 0.5719 & 0.0094 & -0.8373 & 0.0106 \\
2455985.59685 & 0.5689 & 0.0094 & -0.8324 & 0.0106 \\
2455985.5976 & 0.5737 & 0.0095 & -0.8364 & 0.0106 \\
2455985.65429 & 0.5692 & 0.0097 & -0.8359 & 0.0108 \\
2455985.65504 & 0.5728 & 0.0096 & -0.8429 & 0.0108 \\
2455985.65579 & 0.5650 & 0.0096 & -0.8324 & 0.0108 \\
2455985.65653 & 0.5717 & 0.0096 & -0.8364 & 0.0108 \\
2455985.65728 & 0.5826 & 0.0096 & -0.8414 & 0.0107 \\
2455985.65803 & 0.5717 & 0.0096 & -0.8399 & 0.0108 \\
2455985.65877 & 0.5748 & 0.0097 & -0.8332 & 0.0109 \\
2455985.65952 & 0.5775 & 0.0099 & -0.8316 & 0.0110 \\
2455985.66027 & 0.5691 & 0.0096 & -0.8367 & 0.0108 \\
2455985.73744 & 0.5713 & 0.0104 & -0.8211 & 0.0117 \\
2455985.73819 & 0.5724 & 0.0104 & -0.8420 & 0.0118 \\
2455985.73969 & 0.5770 & 0.0105 & -0.8421 & 0.0119 \\
2455985.74043 & 0.5751 & 0.0105 & -0.8431 & 0.0119 \\
2455985.74118 & 0.5686 & 0.0104 & -0.8274 & 0.0118 \\
2455985.74194 & 0.5684 & 0.0104 & -0.8327 & 0.0118 \\
2455985.74268 & 0.5679 & 0.0105 & -0.8442 & 0.0120 \\
2455985.74343 & 0.5640 & 0.0106 & -0.8239 & 0.0120 \\
2455985.74418 & 0.5681 & 0.0105 & -0.8259 & 0.0119 \\
2455985.74493 & 0.5693 & 0.0105 & -0.8253 & 0.0119 \\
2455985.74567 & 0.5657 & 0.0108 & -0.8250 & 0.0122 \\
2455985.74643 & 0.5753 & 0.0107 & -0.8274 & 0.0121 \\
2455986.58227 & 0.5807 & 0.0097 & -0.8349 & 0.0108 \\
2455985.65354 & 0.5764 & 0.0096 & -0.8385 & 0.0108 \\
2455985.73894 & 0.5725 & 0.0105 & -0.8389 & 0.0119 \\
2455986.58303 & 0.5694 & 0.0097 & -0.8385 & 0.0109 \\
2455986.72148 & 0.9822 & 0.0122 & -0.8329 & 0.0113 \\
2455987.5392 & 0.5739 & 0.0097 & -0.8353 & 0.0109 \\
2455987.6327 & 0.5811 & 0.0101 & -0.8334 & 0.0114 \\
2455987.70692 & 0.5785 & 0.0100 & -0.8384 & 0.0113 \\
2455988.54165 & 0.6643 & 0.0104 & -0.8522 & 0.0114 \\
2455988.57402 & 0.8301 & 0.0111 & -0.8365 & 0.0111 \\
2455988.58299 & 0.8984 & 0.0117 & -0.8388 & 0.0114 \\
2455988.59199 & 0.9514 & 0.0120 & -0.8398 & 0.0113 \\
2455986.58377 & 0.5814 & 0.0097 & -0.8322 & 0.0109 \\
2455986.58451 & 0.5726 & 0.0098 & -0.8450 & 0.0110 \\
2455986.58525 & 0.5839 & 0.0098 & -0.8420 & 0.0109 \\
2455986.586 & 0.5765 & 0.0097 & -0.8352 & 0.0109 \\
2455986.58674 & 0.5801 & 0.0097 & -0.8428 & 0.0109 \\
2455986.58748 & 0.5812 & 0.0098 & -0.8313 & 0.0110 \\
2455986.58822 & 0.5759 & 0.0097 & -0.8404 & 0.0109 \\
2455986.58896 & 0.5687 & 0.0097 & -0.8333 & 0.0108 \\
2455986.71925 & 0.9605 & 0.0121 & -0.8315 & 0.0113 \\
2455986.72 & 0.9752 & 0.0121 & -0.8362 & 0.0113 \\
2455986.72074 & 0.9793 & 0.0123 & -0.8239 & 0.0114 \\
2455986.72222 & 0.9884 & 0.0122 & -0.8310 & 0.0113 \\
2455986.72296 & 0.9914 & 0.0121 & -0.8373 & 0.0112 \\
2455986.72371 & 0.9958 & 0.0125 & -0.8351 & 0.0115 \\
2455986.72445 & 0.9939 & 0.0123 & -0.8359 & 0.0113 \\
2455986.72519 & 0.9971 & 0.0125 & -0.8246 & 0.0114 \\
2455986.72594 & 0.9935 & 0.0125 & -0.8321 & 0.0115 \\
2455987.53544 & 0.5667 & 0.0095 & -0.8381 & 0.0106 \\
2455987.53619 & 0.5664 & 0.0101 & -0.8370 & 0.0114 \\
2455987.53694 & 0.5721 & 0.0099 & -0.8353 & 0.0111 \\
2455987.53769 & 0.5666 & 0.0097 & -0.8429 & 0.0110 \\
2455987.53845 & 0.5645 & 0.0096 & -0.8368 & 0.0108 \\
2455987.53994 & 0.5650 & 0.0098 & -0.8316 & 0.0111 \\
2455987.54068 & 0.5714 & 0.0098 & -0.8317 & 0.0110 \\
2455987.54143 & 0.5667 & 0.0099 & -0.8374 & 0.0112 \\
2455987.54218 & 0.5625 & 0.0097 & -0.8411 & 0.0110 \\
2455987.59402 & 0.5721 & 0.0098 & -0.8288 & 0.0110 \\
2455987.59476 & 0.5732 & 0.0096 & -0.8409 & 0.0108 \\
2455987.59551 & 0.5734 & 0.0096 & -0.8520 & 0.0109 \\
2455987.59626 & 0.5777 & 0.0096 & -0.8375 & 0.0107 \\
2455987.597 & 0.5696 & 0.0100 & -0.8352 & 0.0113 \\
2455987.59774 & 0.5757 & 0.0097 & -0.8354 & 0.0109 \\
2455987.59848 & 0.5740 & 0.0099 & -0.8420 & 0.0111 \\
2455987.63345 & 0.5684 & 0.0101 & -0.8348 & 0.0115 \\
2455987.63419 & 0.5781 & 0.0098 & -0.8360 & 0.0110 \\
2455987.63493 & 0.5711 & 0.0100 & -0.8312 & 0.0112 \\
2455987.63569 & 0.5727 & 0.0098 & -0.8361 & 0.0110 \\
2455987.63643 & 0.5722 & 0.0100 & -0.8330 & 0.0113 \\
2455987.63718 & 0.5757 & 0.0101 & -0.8349 & 0.0114 \\
2455987.63792 & 0.5743 & 0.0102 & -0.8406 & 0.0116 \\
2455987.63867 & 0.5781 & 0.0099 & -0.8345 & 0.0111 \\
2455987.63941 & 0.5650 & 0.0101 & -0.8396 & 0.0114 \\
2455987.70542 & 0.5812 & 0.0100 & -0.8414 & 0.0113 \\
2455987.70617 & 0.5784 & 0.0101 & -0.8384 & 0.0115 \\
2455987.70766 & 0.5738 & 0.0100 & -0.8261 & 0.0112 \\
2455987.70842 & 0.5735 & 0.0097 & -0.8310 & 0.0109 \\
2455987.70917 & 0.5814 & 0.0103 & -0.8449 & 0.0116 \\
2455987.70991 & 0.5775 & 0.0100 & -0.8481 & 0.0114 \\
2455987.71066 & 0.5773 & 0.0100 & -0.8385 & 0.0112 \\
2455987.7114 & 0.5765 & 0.0097 & -0.8335 & 0.0109 \\
2455987.71214 & 0.5782 & 0.0099 & -0.8379 & 0.0111 \\
2455988.53867 & 0.6547 & 0.0106 & -0.8476 & 0.0116 \\
2455988.53941 & 0.6547 & 0.0103 & -0.8458 & 0.0113 \\
2455988.54016 & 0.6631 & 0.0104 & -0.8417 & 0.0113 \\
2455988.54091 & 0.6660 & 0.0106 & -0.8382 & 0.0115 \\
2455988.5424 & 0.6798 & 0.0104 & -0.8484 & 0.0112 \\
2455988.54315 & 0.6695 & 0.0107 & -0.8450 & 0.0116 \\
2455988.5439 & 0.6864 & 0.0104 & -0.8379 & 0.0112 \\
2455988.54465 & 0.6835 & 0.0106 & -0.8434 & 0.0115 \\
2455988.54539 & 0.6828 & 0.0104 & -0.8453 & 0.0113 \\
2455988.56957 & 0.8160 & 0.0117 & -0.8490 & 0.0120 \\
2455988.57032 & 0.8186 & 0.0107 & -0.8445 & 0.0108 \\
2455988.57106 & 0.8228 & 0.0110 & -0.8431 & 0.0111 \\
2455988.5718 & 0.8187 & 0.0116 & -0.8410 & 0.0118 \\
2455988.57254 & 0.8360 & 0.0117 & -0.8383 & 0.0117 \\
2455988.57328 & 0.8364 & 0.0117 & -0.8444 & 0.0117 \\
2455988.57476 & 0.8365 & 0.0116 & -0.8339 & 0.0116 \\
2455988.57551 & 0.8469 & 0.0114 & -0.8400 & 0.0114 \\
2455988.57626 & 0.8507 & 0.0114 & -0.8440 & 0.0113 \\
2455988.57701 & 0.8635 & 0.0119 & -0.8413 & 0.0118 \\
2455988.57775 & 0.8630 & 0.0115 & -0.8364 & 0.0114 \\
2455988.57849 & 0.8674 & 0.0118 & -0.8339 & 0.0116 \\
2455988.57925 & 0.8819 & 0.0119 & -0.8435 & 0.0117 \\
2455988.58 & 0.8754 & 0.0120 & -0.8366 & 0.0118 \\
2455988.58075 & 0.8845 & 0.0114 & -0.8397 & 0.0112 \\
2455988.5815 & 0.8897 & 0.0119 & -0.8379 & 0.0116 \\
2455988.58225 & 0.8915 & 0.0114 & -0.8461 & 0.0111 \\
2455988.58375 & 0.9088 & 0.0117 & -0.8461 & 0.0114 \\
2455988.58449 & 0.8968 & 0.0121 & -0.8400 & 0.0118 \\
2455988.58524 & 0.9053 & 0.0116 & -0.8418 & 0.0112 \\
2455988.58598 & 0.9231 & 0.0115 & -0.8492 & 0.0111 \\
2455988.58673 & 0.9260 & 0.0116 & -0.8498 & 0.0112 \\
2455988.58748 & 0.9232 & 0.0119 & -0.8334 & 0.0113 \\
2455988.58824 & 0.9340 & 0.0114 & -0.8422 & 0.0109 \\
2455988.58899 & 0.9290 & 0.0118 & -0.8461 & 0.0113 \\
2455988.58974 & 0.9335 & 0.0120 & -0.8386 & 0.0114 \\
2455988.59049 & 0.9465 & 0.0117 & -0.8407 & 0.0111 \\
2455988.59123 & 0.9413 & 0.0118 & -0.8362 & 0.0112 \\
2455988.59274 & 0.9582 & 0.0119 & -0.8334 & 0.0111 \\
2455988.59349 & 0.9627 & 0.0121 & -0.8313 & 0.0113 \\
2455988.59424 & 0.9686 & 0.0119 & -0.8411 & 0.0111 \\
2455988.595 & 0.9680 & 0.0120 & -0.8402 & 0.0112 \\
2455988.59574 & 0.9760 & 0.0120 & -0.8352 & 0.0112 \\
2455988.59649 & 0.9659 & 0.0120 & -0.8321 & 0.0112 \\
2455988.59724 & 0.9841 & 0.0121 & -0.8360 & 0.0112 \\
2455988.59798 & 0.9852 & 0.0118 & -0.8386 & 0.0110 \\
2455988.59872 & 0.9928 & 0.0122 & -0.8389 & 0.0112 \\
2455988.6009 & 0.9998 & 0.0121 & -0.8382 & 0.0112 \\
2455988.60164 & 1.0018 & 0.0123 & -0.8330 & 0.0112 \\
2455988.60312 & 1.0115 & 0.0122 & -0.8350 & 0.0112 \\
2455988.60386 & 1.0013 & 0.0123 & -0.8362 & 0.0113 \\
2455988.6046 & 1.0139 & 0.0118 & -0.8383 & 0.0108 \\
2455988.60535 & 1.0168 & 0.0121 & -0.8391 & 0.0110 \\
2455988.60611 & 1.0370 & 0.0121 & -0.8436 & 0.0110 \\
2455988.60685 & 1.0245 & 0.0120 & -0.8302 & 0.0109 \\
2455988.60759 & 1.0293 & 0.0120 & -0.8337 & 0.0109 \\
2455988.60833 & 1.0297 & 0.0126 & -0.8482 & 0.0114 \\
2455988.60908 & 1.0261 & 0.0119 & -0.8306 & 0.0108 \\
2455988.60983 & 1.0321 & 0.0120 & -0.8320 & 0.0109 \\
2455988.61058 & 1.0241 & 0.0120 & -0.8427 & 0.0109 \\
2455988.61209 & 1.0281 & 0.0123 & -0.8340 & 0.0111 \\
2455988.61284 & 1.0362 & 0.0123 & -0.8430 & 0.0112 \\
2455988.61358 & 1.0257 & 0.0123 & -0.8343 & 0.0111 \\
2455988.61432 & 1.0346 & 0.0125 & -0.8369 & 0.0113 \\
2455988.61508 & 1.0428 & 0.0122 & -0.8304 & 0.0110 \\
2455988.61583 & 1.0258 & 0.0121 & -0.8459 & 0.0111 \\
2455988.61658 & 1.0376 & 0.0122 & -0.8296 & 0.0110 \\
2455988.61732 & 1.0318 & 0.0122 & -0.8323 & 0.0110 \\
2455988.61806 & 1.0312 & 0.0124 & -0.8493 & 0.0113 \\
2455988.6188 & 1.0313 & 0.0127 & -0.8363 & 0.0114 \\
2455988.61955 & 1.0265 & 0.0125 & -0.8377 & 0.0114 \\
2455988.62106 & 1.0309 & 0.0122 & -0.8353 & 0.0110 \\
2455988.62181 & 1.0331 & 0.0125 & -0.8335 & 0.0113 \\
2455988.62256 & 1.0366 & 0.0122 & -0.8356 & 0.0110 \\
2455988.62348 & 1.0287 & 0.0127 & -0.8400 & 0.0115 \\
2455988.62422 & 1.0266 & 0.0122 & -0.8461 & 0.0111 \\
2455988.62497 & 1.0141 & 0.0128 & -0.8470 & 0.0117 \\
2455988.62594 & 1.0135 & 0.0124 & -0.8488 & 0.0114 \\
2455988.62668 & 1.0227 & 0.0124 & -0.8400 & 0.0113 \\
2455988.62744 & 1.0113 & 0.0117 & -0.8396 & 0.0108 \\
2455988.62818 & 0.9986 & 0.0118 & -0.8412 & 0.0109 \\
2455988.62893 & 0.9919 & 0.0118 & -0.8351 & 0.0109 \\
2455988.63042 & 0.9961 & 0.0122 & -0.8353 & 0.0112 \\
2455988.63117 & 0.9903 & 0.0122 & -0.8344 & 0.0112 \\
2455988.63191 & 0.9931 & 0.0121 & -0.8472 & 0.0112 \\
2455988.63266 & 0.9832 & 0.0120 & -0.8351 & 0.0111 \\
2455988.63348 & 0.9777 & 0.0123 & -0.8369 & 0.0114 \\
2455988.63423 & 0.9754 & 0.0123 & -0.8342 & 0.0114 \\
2455988.63498 & 0.9750 & 0.0121 & -0.8355 & 0.0112 \\
2455988.63573 & 0.9792 & 0.0123 & -0.8392 & 0.0114 \\
2455988.63649 & 0.9649 & 0.0119 & -0.8363 & 0.0111 \\
2455988.6616 & 0.8144 & 0.0113 & -0.8485 & 0.0115 \\
2455988.66235 & 0.8010 & 0.0113 & -0.8375 & 0.0115 \\
2455988.66385 & 0.7942 & 0.0114 & -0.8295 & 0.0116 \\
2455988.6646 & 0.7941 & 0.0114 & -0.8345 & 0.0117 \\
2455988.69095 & 0.6588 & 0.0110 & -0.8384 & 0.0120 \\
2455988.6917 & 0.6469 & 0.0106 & -0.8310 & 0.0116 \\
2455988.69246 & 0.6524 & 0.0109 & -0.8274 & 0.0119 \\
2455988.72766 & 0.5678 & 0.0109 & -0.8375 & 0.0124 \\
2455988.7284 & 0.5724 & 0.0116 & -0.8296 & 0.0132 \\
2455988.72916 & 0.5686 & 0.0109 & -0.8263 & 0.0123 \\
2455988.75337 & 0.5721 & 0.0111 & -0.8361 & 0.0126 \\
2455988.75412 & 0.5677 & 0.0105 & -0.8244 & 0.0119 \\
2455988.75487 & 0.5593 & 0.0113 & -0.8249 & 0.0128 \\
2455989.51787 & 0.5752 & 0.0102 & -0.8356 & 0.0115 \\
2455989.51862 & 0.5709 & 0.0105 & -0.8445 & 0.0119 \\
2455989.54275 & 0.5790 & 0.0098 & -0.8452 & 0.0111 \\
2455989.54348 & 0.5685 & 0.0100 & -0.8417 & 0.0114 \\
2455989.57261 & 0.5781 & 0.0104 & -0.8446 & 0.0118 \\
2455989.57336 & 0.5782 & 0.0100 & -0.8412 & 0.0113 \\
2455989.60177 & 0.5794 & 0.0103 & -0.8339 & 0.0116 \\
2455989.60252 & 0.5747 & 0.0102 & -0.8394 & 0.0116 \\
2455989.61017 & 0.5839 & 0.0097 & -0.8459 & 0.0108 \\
2455989.61092 & 0.5887 & 0.0096 & -0.8485 & 0.0108 \\
2455989.62016 & 0.5989 & 0.0103 & -0.8480 & 0.0116 \\
2455989.62165 & 0.6034 & 0.0101 & -0.8507 & 0.0113 \\
2455989.63729 & 0.6328 & 0.0106 & -0.8439 & 0.0117 \\
2455989.63804 & 0.6375 & 0.0105 & -0.8384 & 0.0116 \\
2455989.63879 & 0.6428 & 0.0104 & -0.8399 & 0.0114 \\
2455989.63954 & 0.6436 & 0.0105 & -0.8450 & 0.0116 \\
2455989.64029 & 0.6461 & 0.0104 & -0.8390 & 0.0114 \\
2455989.65347 & 0.6960 & 0.0103 & -0.8379 & 0.0110 \\
2455989.65422 & 0.6983 & 0.0107 & -0.8427 & 0.0115 \\
2455989.65497 & 0.7058 & 0.0111 & -0.8405 & 0.0119 \\
2455989.65572 & 0.6995 & 0.0106 & -0.8409 & 0.0113 \\
2455989.65647 & 0.7067 & 0.0110 & -0.8399 & 0.0117 \\
2455989.68327 & 0.8495 & 0.0119 & -0.8427 & 0.0119 \\
2455989.68401 & 0.8488 & 0.0119 & -0.8276 & 0.0118 \\
2455989.70833 & 0.9829 & 0.0141 & -0.8415 & 0.0130 \\
2455989.70908 & 0.9880 & 0.0134 & -0.8336 & 0.0124 \\
2455989.70982 & 0.9855 & 0.0138 & -0.8456 & 0.0128 \\
2455989.73451 & 1.0151 & 0.0143 & -0.8275 & 0.0129 \\
2455989.73526 & 0.9980 & 0.0148 & -0.8435 & 0.0136 \\
2455989.736 & 0.9982 & 0.0141 & -0.8286 & 0.0129 \\
2455989.75485 & 0.9228 & 0.0127 & -0.8422 & 0.0122 \\
2455989.7556 & 0.9104 & 0.0126 & -0.8329 & 0.0121 \\
2455989.75636 & 0.9071 & 0.0137 & -0.8382 & 0.0132 \\
2455989.75786 & 0.8750 & 0.0159 & -0.8262 & 0.0155 \\
2455989.76794 & 0.8230 & 0.0132 & -0.8482 & 0.0134 \\
2455989.76869 & 0.8228 & 0.0127 & -0.8396 & 0.0128 \\
2455989.76943 & 0.8236 & 0.0123 & -0.8454 & 0.0125 \\
2455989.77019 & 0.8153 & 0.0132 & -0.8392 & 0.0134 \\
2455989.77093 & 0.8005 & 0.0134 & -0.8474 & 0.0137 \\
2455989.77167 & 0.8024 & 0.0133 & -0.8446 & 0.0136 \\
2455989.77242 & 0.7918 & 0.0131 & -0.8390 & 0.0135 \\
2455989.77316 & 0.7845 & 0.0126 & -0.8300 & 0.0129 \\
2455989.7739 & 0.7977 & 0.0129 & -0.8312 & 0.0131 \\
2455989.77464 & 0.7882 & 0.0128 & -0.8342 & 0.0131 \\
2455989.77614 & 0.7744 & 0.0135 & -0.8475 & 0.0140 \\
2456287.68142 & 0.5630 & 0.0099 & -0.8374 & 0.0113 \\
2456287.68225 & 0.5635 & 0.0099 & -0.8319 & 0.0113 \\
2456287.68306 & 0.5591 & 0.0098 & -0.8403 & 0.0113 \\
2456287.68389 & 0.5793 & 0.0100 & -0.8527 & 0.0114 \\
2456287.68472 & 0.5591 & 0.0101 & -0.8475 & 0.0116 \\
2456287.74799 & 0.5600 & 0.0100 & -0.8466 & 0.0116 \\
2456287.74883 & 0.5651 & 0.0100 & -0.8384 & 0.0114 \\
2456287.74967 & 0.5635 & 0.0098 & -0.8461 & 0.0112 \\
2456287.75051 & 0.5743 & 0.0099 & -0.8332 & 0.0112 \\
2456287.75135 & 0.5595 & 0.0098 & -0.8371 & 0.0112 \\
2456287.76498 & 0.5675 & 0.0097 & -0.8365 & 0.0110 \\
2456287.76582 & 0.5589 & 0.0097 & -0.8299 & 0.0110 \\
2456287.76665 & 0.5790 & 0.0097 & -0.8367 & 0.0109 \\
2456287.76747 & 0.5707 & 0.0097 & -0.8519 & 0.0112 \\
2456287.76822 & 0.5696 & 0.0097 & -0.8459 & 0.0111 \\
2456287.76903 & 0.5699 & 0.0098 & -0.8433 & 0.0112 \\
2456287.76988 & 0.5692 & 0.0097 & -0.8410 & 0.0111 \\
2456287.77072 & 0.5849 & 0.0097 & -0.8433 & 0.0111 \\
2456287.77153 & 0.5670 & 0.0098 & -0.8483 & 0.0112 \\
2456287.77228 & 0.5632 & 0.0098 & -0.8477 & 0.0113 \\
2456287.77311 & 0.5761 & 0.0099 & -0.8416 & 0.0112 \\
2456287.77479 & 0.5838 & 0.0099 & -0.8495 & 0.0113 \\
2456287.77561 & 0.5715 & 0.0099 & -0.8385 & 0.0113 \\
2456287.77635 & 0.5639 & 0.0098 & -0.8516 & 0.0113 \\
2456287.77717 & 0.5604 & 0.0099 & -0.8489 & 0.0114 \\
2456287.77801 & 0.5596 & 0.0098 & -0.8429 & 0.0112 \\
2456287.77888 & 0.5841 & 0.0099 & -0.8445 & 0.0113 \\
2456287.77971 & 0.5674 & 0.0098 & -0.8464 & 0.0113 \\
2456287.78046 & 0.5613 & 0.0098 & -0.8516 & 0.0113 \\
2456287.78127 & 0.5664 & 0.0100 & -0.8460 & 0.0114 \\
2456287.78208 & 0.5618 & 0.0098 & -0.8334 & 0.0112 \\
2456287.78293 & 0.5812 & 0.0098 & -0.8532 & 0.0111 \\
2455988.61134 & 1.0341 & 0.0121 & -0.8420 & 0.0110 \\
2455988.62031 & 1.0308 & 0.0130 & -0.8366 & 0.0117 \\
2455988.62968 & 0.9937 & 0.0125 & -0.8345 & 0.0115 \\
2455988.66309 & 0.7890 & 0.0115 & -0.8357 & 0.0117 \\
2455989.51712 & 0.5720 & 0.0104 & -0.8446 & 0.0119 \\
2455989.62091 & 0.5908 & 0.0102 & -0.8478 & 0.0115 \\
2455989.68252 & 0.8334 & 0.0120 & -0.8376 & 0.0120 \\
2455989.75711 & 0.8883 & 0.0139 & -0.8371 & 0.0135 \\
2455989.77538 & 0.7930 & 0.0129 & -0.8487 & 0.0132 \\
2456287.76417 & 0.5605 & 0.0096 & -0.8351 & 0.0110 \\
2456287.77395 & 0.5585 & 0.0098 & -0.8356 & 0.0112 \\
2456287.78449 & 0.5615 & 0.0098 & -0.8460 & 0.0113 \\
2456287.78532 & 0.5753 & 0.0098 & -0.8450 & 0.0112 \\
2456287.78616 & 0.5658 & 0.0097 & -0.8385 & 0.0111 \\
2456287.787 & 0.5874 & 0.0097 & -0.8319 & 0.0109 \\
2456287.78784 & 0.5665 & 0.0097 & -0.8523 & 0.0112 \\
2456287.82224 & 0.5570 & 0.0096 & -0.8531 & 0.0111 \\
2456287.82305 & 0.5625 & 0.0097 & -0.8348 & 0.0110 \\
2456287.82387 & 0.5642 & 0.0096 & -0.8460 & 0.0110 \\
2456287.82471 & 0.5853 & 0.0098 & -0.8451 & 0.0111 \\
2456287.82553 & 0.5677 & 0.0098 & -0.8417 & 0.0112 \\
2456287.84051 & 0.5634 & 0.0096 & -0.8460 & 0.0110 \\
2456287.84215 & 0.5683 & 0.0095 & -0.8450 & 0.0108 \\
2456287.84299 & 0.5829 & 0.0095 & -0.8520 & 0.0108 \\
2456287.8438 & 0.5680 & 0.0095 & -0.8415 & 0.0108 \\
2456287.8479 & 0.5657 & 0.0092 & -0.8495 & 0.0105 \\
2456287.84872 & 0.5654 & 0.0093 & -0.8278 & 0.0105 \\
2456287.84953 & 0.5622 & 0.0092 & -0.8358 & 0.0104 \\
2456287.85038 & 0.5694 & 0.0092 & -0.8336 & 0.0104 \\
2456287.85119 & 0.5725 & 0.0092 & -0.8455 & 0.0104 \\
2456307.59668 & 0.7436 & 0.0096 & -0.8445 & 0.0100 \\
2456307.59742 & 0.7451 & 0.0096 & -0.8494 & 0.0100 \\
2456307.59818 & 0.7500 & 0.0096 & -0.8528 & 0.0101 \\
2456307.59968 & 0.7659 & 0.0096 & -0.8534 & 0.0100 \\
2456307.6583 & 1.0451 & 0.0111 & -0.8395 & 0.0100 \\
2456307.65904 & 1.0397 & 0.0109 & -0.8496 & 0.0100 \\
2456307.6598 & 1.0432 & 0.0110 & -0.8381 & 0.0100 \\
2456307.66054 & 1.0400 & 0.0110 & -0.8395 & 0.0100 \\
2456307.66129 & 1.0417 & 0.0111 & -0.8382 & 0.0101 \\
2456307.69371 & 0.8384 & 0.0100 & -0.8350 & 0.0100 \\
2456307.69446 & 0.8330 & 0.0100 & -0.8312 & 0.0100 \\
2456307.69521 & 0.8303 & 0.0100 & -0.8305 & 0.0100 \\
2456307.69595 & 0.8258 & 0.0100 & -0.8316 & 0.0100 \\
2456307.69669 & 0.8237 & 0.0099 & -0.8324 & 0.0100 \\
2456307.69818 & 0.8122 & 0.0099 & -0.8415 & 0.0100 \\
2456307.83933 & 0.5675 & 0.0096 & -0.8330 & 0.0108 \\
2456307.84007 & 0.5787 & 0.0095 & -0.8346 & 0.0107 \\
2456307.84081 & 0.5747 & 0.0095 & -0.8228 & 0.0106 \\
2456307.84157 & 0.5814 & 0.0095 & -0.8311 & 0.0106 \\
2456307.84232 & 0.5681 & 0.0095 & -0.8331 & 0.0107 \\
2456307.84306 & 0.5753 & 0.0095 & -0.8297 & 0.0106 \\
2456307.8438 & 0.5672 & 0.0095 & -0.8282 & 0.0107 \\
2456308.52119 & 0.5740 & 0.0487 & -0.8766 & 0.0606 \\
2456308.52194 & 0.5612 & 0.0433 & -0.8657 & 0.0538 \\
2456308.52269 & 0.5665 & 0.0395 & -0.8486 & 0.0483 \\
2456308.52419 & 0.5656 & 0.0325 & -0.8570 & 0.0399 \\
2456308.52494 & 0.5442 & 0.0294 & -0.8335 & 0.0360 \\
2456308.5257 & 0.5490 & 0.0268 & -0.8511 & 0.0330 \\
2456308.52645 & 0.5844 & 0.0254 & -0.8461 & 0.0305 \\
2456308.52719 & 0.5629 & 0.0239 & -0.8383 & 0.0288 \\
2456308.52794 & 0.5682 & 0.0215 & -0.8284 & 0.0257 \\
2456308.52868 & 0.5571 & 0.0199 & -0.8447 & 0.0241 \\
2456308.52944 & 0.5622 & 0.0182 & -0.8477 & 0.0220 \\
2456308.6016 & 0.5699 & 0.0090 & -0.8434 & 0.0101 \\
2456308.60234 & 0.5844 & 0.0090 & -0.8445 & 0.0101 \\
2456308.60309 & 0.5766 & 0.0090 & -0.8479 & 0.0101 \\
2456308.6046 & 0.5690 & 0.0090 & -0.8401 & 0.0101 \\
2456308.60535 & 0.5723 & 0.0090 & -0.8460 & 0.0102 \\
2456308.6061 & 0.5710 & 0.0090 & -0.8453 & 0.0102 \\
2456308.84739 & 0.6321 & 0.0100 & -0.8360 & 0.0110 \\
2456308.84814 & 0.6289 & 0.0099 & -0.8299 & 0.0109 \\
2456308.84888 & 0.6294 & 0.0100 & -0.8394 & 0.0110 \\
2456308.84962 & 0.6288 & 0.0100 & -0.8415 & 0.0110 \\
2456309.5247 & 0.9573 & 0.0900 & -0.8476 & 0.0828 \\
2456309.52545 & 1.0254 & 0.0532 & -0.8544 & 0.0467 \\
2456309.5262 & 1.0521 & 0.0471 & -0.8398 & 0.0401 \\
2456309.52694 & 1.0537 & 0.0447 & -0.8335 & 0.0378 \\
2456309.52845 & 1.0124 & 0.0547 & -0.8570 & 0.0487 \\
2456309.52919 & 1.0300 & 0.0999 & -0.7846 & 0.0828 \\
2456309.52993 & 1.0733 & 0.1405 & -0.8457 & 0.1178 \\
2456309.53067 & 0.9993 & 0.0553 & -0.8558 & 0.0497 \\
2456309.53142 & 1.0562 & 0.0729 & -0.7958 & 0.0600 \\
2456309.59252 & 0.7046 & 0.0105 & -0.8238 & 0.0111 \\
2456309.59326 & 0.7056 & 0.0109 & -0.8406 & 0.0117 \\
2456309.59401 & 0.7025 & 0.0112 & -0.8346 & 0.0119 \\
2456309.59475 & 0.6980 & 0.0103 & -0.8354 & 0.0110 \\
2456309.62351 & 0.5919 & 0.0094 & -0.8391 & 0.0105 \\
2456309.62426 & 0.5938 & 0.0093 & -0.8528 & 0.0105 \\
2456309.62575 & 0.5888 & 0.0092 & -0.8401 & 0.0102 \\
2456309.75394 & 0.5856 & 0.0125 & -0.8372 & 0.0141 \\
2456309.75468 & 0.5862 & 0.0148 & -0.8339 & 0.0168 \\
2456309.75544 & 0.5914 & 0.0145 & -0.8499 & 0.0165 \\
2456309.75618 & 0.5809 & 0.0131 & -0.8367 & 0.0149 \\
2456310.86793 & 0.5543 & 0.0114 & -0.8219 & 0.0131 \\
2456310.86868 & 0.5524 & 0.0113 & -0.8137 & 0.0129 \\
2456310.86942 & 0.5651 & 0.0113 & -0.8245 & 0.0129 \\
2456310.87017 & 0.5667 & 0.0116 & -0.8357 & 0.0133 \\
2456310.87092 & 0.5632 & 0.0116 & -0.8208 & 0.0133 \\
2456310.87166 & 0.5610 & 0.0121 & -0.8297 & 0.0139 \\
2456311.52929 & 0.5658 & 0.0192 & -0.8109 & 0.0224 \\
2456311.53003 & 0.5785 & 0.0184 & -0.8416 & 0.0218 \\
2456311.53078 & 0.5889 & 0.0201 & -0.8406 & 0.0236 \\
2456311.53152 & 0.5712 & 0.0171 & -0.8557 & 0.0204 \\
2456311.53227 & 0.5750 & 0.0157 & -0.8259 & 0.0183 \\
2456311.53301 & 0.5772 & 0.0146 & -0.8474 & 0.0172 \\
2456311.53377 & 0.5759 & 0.0147 & -0.8348 & 0.0171 \\
2456311.53451 & 0.5813 & 0.0140 & -0.8386 & 0.0163 \\
2456311.53526 & 0.5676 & 0.0124 & -0.8319 & 0.0143 \\
2456311.53601 & 0.5806 & 0.0117 & -0.8412 & 0.0134 \\
2456311.54614 & 0.5600 & 0.0125 & -0.8173 & 0.0142 \\
2456311.54763 & 0.5667 & 0.0126 & -0.8177 & 0.0143 \\
2456311.58593 & 0.5560 & 0.0105 & -0.8285 & 0.0119 \\
2456311.58667 & 0.5775 & 0.0118 & -0.8409 & 0.0134 \\
2456311.58741 & 0.5694 & 0.0203 & -0.8303 & 0.0236 \\
2456311.58816 & 0.5622 & 0.0181 & -0.8212 & 0.0210 \\
2456311.58937 & 0.5668 & 0.0108 & -0.8401 & 0.0123 \\
2456311.59012 & 0.5642 & 0.0098 & -0.8378 & 0.0111 \\
2456311.59087 & 0.5684 & 0.0099 & -0.8284 & 0.0112 \\
2456311.59161 & 0.5707 & 0.0100 & -0.8308 & 0.0112 \\
2456311.88389 & 0.5399 & 0.0158 & -0.8087 & 0.0185 \\
2456311.88463 & 0.5592 & 0.0172 & -0.7991 & 0.0199 \\
2456446.43771 & 0.6803 & 0.0545 & -0.8293 & 0.0607 \\
2456446.43846 & 0.6756 & 0.0491 & -0.8785 & 0.0570 \\
2456446.43921 & 0.6369 & 0.0439 & -0.8563 & 0.0515 \\
2456446.43996 & 0.6716 & 0.0413 & -0.7915 & 0.0450 \\
2456446.44072 & 0.6339 & 0.0357 & -0.8548 & 0.0418 \\
2456446.44146 & 0.6536 & 0.0330 & -0.8487 & 0.0380 \\
2456446.4422 & 0.6334 & 0.0297 & -0.8573 & 0.0348 \\
2456446.44295 & 0.6365 & 0.0274 & -0.8265 & 0.0312 \\
2456446.4437 & 0.6286 & 0.0247 & -0.8420 & 0.0286 \\
2456446.44444 & 0.6384 & 0.0232 & -0.8515 & 0.0268 \\
2456446.4452 & 0.6434 & 0.0213 & -0.8298 & 0.0241 \\
2456446.44668 & 0.6370 & 0.0182 & -0.8505 & 0.0210 \\
2456446.46542 & 0.7018 & 0.0108 & -0.8378 & 0.0115 \\
2456446.46617 & 0.6986 & 0.0107 & -0.8276 & 0.0114 \\
2456446.46691 & 0.7084 & 0.0107 & -0.8359 & 0.0114 \\
2456446.46765 & 0.7136 & 0.0107 & -0.8364 & 0.0114 \\
2456446.4684 & 0.7152 & 0.0108 & -0.8348 & 0.0115 \\
2456446.50272 & 0.9036 & 0.0124 & -0.8155 & 0.0118 \\
2456446.50347 & 0.9046 & 0.0124 & -0.8131 & 0.0118 \\
2456446.50421 & 0.9106 & 0.0124 & -0.8252 & 0.0119 \\
2456446.50495 & 0.9213 & 0.0125 & -0.8257 & 0.0119 \\
2456446.50569 & 0.9210 & 0.0126 & -0.8141 & 0.0119 \\
2456287.84133 & 0.5620 & 0.0095 & -0.8301 & 0.0108 \\
2456307.59893 & 0.7493 & 0.0096 & -0.8533 & 0.0101 \\
2456307.69743 & 0.8260 & 0.0099 & -0.8408 & 0.0100 \\
2456308.52344 & 0.5328 & 0.0351 & -0.8682 & 0.0445 \\
2456308.60385 & 0.5737 & 0.0090 & -0.8217 & 0.0100 \\
2456309.52769 & 1.0146 & 0.0467 & -0.8282 & 0.0406 \\
2456309.62501 & 0.5883 & 0.0093 & -0.8422 & 0.0104 \\
2456310.8724 & 0.5525 & 0.0123 & -0.8185 & 0.0142 \\
2456311.54689 & 0.5700 & 0.0124 & -0.8336 & 0.0141 \\
2456311.88539 & 0.4848 & 0.0169 & -0.8179 & 0.0208 \\
2456446.44594 & 0.6422 & 0.0198 & -0.8592 & 0.0229 \\
2455988.60238 & 1.0015 & 0.0122 & -0.8321 & 0.0112 \\
2456287.78374 & 0.5741 & 0.0098 & -0.8536 & 0.0113 \\
2456447.43757 & 0.6524 & 0.0547 & -0.8311 & 0.0622 \\
2456447.43832 & 0.6131 & 0.0475 & -0.8377 & 0.0558 \\
2456447.43906 & 0.6071 & 0.0422 & -0.8560 & 0.0504 \\
2456447.43982 & 0.6138 & 0.0394 & -0.8531 & 0.0467 \\
2456447.44056 & 0.5984 & 0.0348 & -0.8241 & 0.0408 \\
2456447.4413 & 0.5897 & 0.0318 & -0.8579 & 0.0384 \\
2456447.44205 & 0.5886 & 0.0285 & -0.8922 & 0.0354 \\
2456447.44279 & 0.5778 & 0.0268 & -0.8369 & 0.0321 \\
2456447.44353 & 0.5716 & 0.0246 & -0.8219 & 0.0292 \\
2456447.44427 & 0.5962 & 0.0229 & -0.8596 & 0.0274 \\
2456447.4775 & 0.5541 & 0.0104 & -0.8140 & 0.0118 \\
2456447.47825 & 0.5493 & 0.0104 & -0.8139 & 0.0119 \\
2456447.47976 & 0.5553 & 0.0105 & -0.8232 & 0.0120 \\
2456447.4805 & 0.5433 & 0.0105 & -0.8124 & 0.0120 \\
2456447.50404 & 0.5524 & 0.0108 & -0.8178 & 0.0124 \\
2456447.50479 & 0.5452 & 0.0109 & -0.8069 & 0.0125 \\
2456447.50553 & 0.5406 & 0.0110 & -0.8056 & 0.0126 \\
2456447.50628 & 0.5357 & 0.0110 & -0.8085 & 0.0126 \\
2456447.50702 & 0.5375 & 0.0110 & -0.8009 & 0.0126 \\
2456448.44369 & 0.8709 & 0.0278 & -0.8508 & 0.0274 \\
2456448.44444 & 0.8844 & 0.0258 & -0.8502 & 0.0252 \\
2456448.44519 & 0.8598 & 0.0233 & -0.8500 & 0.0232 \\
2456448.44593 & 0.8785 & 0.0221 & -0.8393 & 0.0215 \\
2456448.48129 & 0.6538 & 0.0107 & -0.8291 & 0.0117 \\
2456448.48203 & 0.6519 & 0.0108 & -0.8264 & 0.0117 \\
2456448.48277 & 0.6496 & 0.0108 & -0.8224 & 0.0118 \\
2456448.48351 & 0.6433 & 0.0108 & -0.8212 & 0.0117 \\
2456448.48425 & 0.6435 & 0.0108 & -0.7829 & 0.0115 \\
2456448.51656 & 0.5727 & 0.0116 & -0.8252 & 0.0133 \\
2456448.51732 & 0.5696 & 0.0118 & -0.8327 & 0.0136 \\
2456448.51807 & 0.5655 & 0.0119 & -0.8086 & 0.0135 \\
2456448.51883 & 0.5690 & 0.0117 & -0.8137 & 0.0133 \\
2456448.51957 & 0.5680 & 0.0117 & -0.8128 & 0.0134 \\
2456449.46411 & 0.7900 & 0.0116 & -0.8341 & 0.0119 \\
2456449.4656 & 0.7884 & 0.0116 & -0.8377 & 0.0119 \\
2456449.46634 & 0.7919 & 0.0118 & -0.8252 & 0.0120 \\
2456449.46708 & 0.7894 & 0.0117 & -0.8238 & 0.0120 \\
2456449.46783 & 0.7940 & 0.0117 & -0.8273 & 0.0120 \\
2456449.46857 & 0.8041 & 0.0118 & -0.8337 & 0.0120 \\
2456449.49085 & 0.9186 & 0.0137 & -0.8178 & 0.0129 \\
2456449.49159 & 0.9196 & 0.0137 & -0.8023 & 0.0129 \\
2456449.49234 & 0.9280 & 0.0137 & -0.8027 & 0.0128 \\
2456449.49309 & 0.9276 & 0.0139 & -0.8224 & 0.0131 \\
2456449.49383 & 0.9183 & 0.0138 & -0.8051 & 0.0129 \\
2456449.49458 & 0.9390 & 0.0139 & -0.8124 & 0.0130 \\
2456449.51859 & 1.0230 & 0.0150 & -0.8228 & 0.0133 \\
2456449.51933 & 1.0261 & 0.0150 & -0.8381 & 0.0134 \\
2456449.52009 & 1.0316 & 0.0155 & -0.8152 & 0.0136 \\
2456449.52084 & 1.0211 & 0.0154 & -0.8153 & 0.0136 \\
2456449.52159 & 1.0252 & 0.0151 & -0.8197 & 0.0134 \\
2456449.52233 & 1.0206 & 0.0151 & -0.8297 & 0.0135 \\
2456449.52307 & 1.0267 & 0.0154 & -0.8123 & 0.0136 \\
2456450.47141 & 0.5348 & 0.0129 & -0.8157 & 0.0150 \\
2456450.47217 & 0.5407 & 0.0134 & -0.8209 & 0.0157 \\
2456450.47292 & 0.5506 & 0.0135 & -0.8338 & 0.0158 \\
2456450.47367 & 0.5401 & 0.0124 & -0.8264 & 0.0145 \\
2456450.47516 & 0.5458 & 0.0117 & -0.8385 & 0.0137 \\
2456450.4759 & 0.5565 & 0.0118 & -0.8396 & 0.0137 \\
2456450.50959 & 0.5649 & 0.0130 & -0.8051 & 0.0148 \\
2456450.51033 & 0.5620 & 0.0129 & -0.8099 & 0.0148 \\
2456450.51109 & 0.5694 & 0.0129 & -0.8169 & 0.0148 \\
2456450.51184 & 0.5502 & 0.0133 & -0.8128 & 0.0154 \\
2456450.51258 & 0.5736 & 0.0130 & -0.8107 & 0.0148 \\
2456450.51333 & 0.5763 & 0.0128 & -0.8185 & 0.0147 \\
2456450.51408 & 0.5646 & 0.0128 & -0.8144 & 0.0147 \\
2456712.53856 & 0.5643 & 0.0093 & -0.8316 & 0.0104 \\
2456712.53931 & 0.5666 & 0.0092 & -0.8328 & 0.0103 \\
2456712.54079 & 0.5604 & 0.0092 & -0.8293 & 0.0103 \\
2456712.54153 & 0.5670 & 0.0092 & -0.8239 & 0.0103 \\
2456712.54228 & 0.5642 & 0.0092 & -0.8233 & 0.0103 \\
2456712.54302 & 0.5682 & 0.0092 & -0.8301 & 0.0103 \\
2456712.54376 & 0.5663 & 0.0092 & -0.8265 & 0.0103 \\
2456712.5445 & 0.5624 & 0.0092 & -0.8252 & 0.0103 \\
2456712.54524 & 0.5646 & 0.0092 & -0.8290 & 0.0103 \\
2456712.54599 & 0.5673 & 0.0092 & -0.8342 & 0.0103 \\
2456712.54673 & 0.5646 & 0.0092 & -0.8311 & 0.0103 \\
2456712.54748 & 0.5683 & 0.0092 & -0.8391 & 0.0103 \\
2456712.54822 & 0.5562 & 0.0092 & -0.8371 & 0.0103 \\
2456712.59976 & 0.5682 & 0.0093 & -0.8390 & 0.0104 \\
2456712.60051 & 0.5726 & 0.0093 & -0.8397 & 0.0104 \\
2456712.60125 & 0.5686 & 0.0093 & -0.8354 & 0.0104 \\
2456712.602 & 0.5680 & 0.0093 & -0.8369 & 0.0104 \\
2456712.63767 & 0.5661 & 0.0094 & -0.8270 & 0.0105 \\
2456712.63841 & 0.5691 & 0.0093 & -0.8419 & 0.0105 \\
2456712.63915 & 0.5592 & 0.0093 & -0.8450 & 0.0106 \\
2456712.70392 & 0.5623 & 0.0103 & -0.8369 & 0.0117 \\
2456712.70466 & 0.5599 & 0.0098 & -0.8381 & 0.0111 \\
2456712.70541 & 0.5653 & 0.0098 & -0.8405 & 0.0111 \\
2456712.70615 & 0.5643 & 0.0103 & -0.8423 & 0.0117 \\
2456713.51054 & 0.5525 & 0.0154 & -0.8353 & 0.0182 \\
2456713.51129 & 0.5589 & 0.0136 & -0.8282 & 0.0158 \\
2456713.51203 & 0.5481 & 0.0135 & -0.8173 & 0.0157 \\
2456713.51278 & 0.5528 & 0.0126 & -0.8215 & 0.0146 \\
2456713.51352 & 0.5574 & 0.0121 & -0.8225 & 0.0140 \\
2456713.51426 & 0.5573 & 0.0116 & -0.8216 & 0.0133 \\
2456713.515 & 0.5602 & 0.0115 & -0.8374 & 0.0133 \\
2456713.51575 & 0.5513 & 0.0110 & -0.8298 & 0.0126 \\
2456713.51649 & 0.5609 & 0.0107 & -0.8292 & 0.0123 \\
2456713.51723 & 0.5499 & 0.0109 & -0.8297 & 0.0125 \\
2456713.64322 & 0.5664 & 0.0099 & -0.8347 & 0.0111 \\
2456713.6447 & 0.5723 & 0.0097 & -0.8372 & 0.0110 \\
2456713.64544 & 0.5729 & 0.0099 & -0.8374 & 0.0111 \\
2456713.64619 & 0.5796 & 0.0098 & -0.8426 & 0.0111 \\
2456713.72604 & 0.5578 & 0.0116 & -0.8295 & 0.0132 \\
2456713.72678 & 0.5720 & 0.0117 & -0.8261 & 0.0132 \\
2456713.72752 & 0.5595 & 0.0119 & -0.8102 & 0.0134 \\
2456713.72826 & 0.5573 & 0.0122 & -0.8294 & 0.0139 \\
2456713.729 & 0.5621 & 0.0119 & -0.8311 & 0.0135 \\
2456714.5201 & 0.5498 & 0.0103 & -0.8402 & 0.0119 \\
2456714.52084 & 0.5511 & 0.0103 & -0.8348 & 0.0118 \\
2456714.52158 & 0.5403 & 0.0104 & -0.8238 & 0.0119 \\
2456714.52306 & 0.5445 & 0.0106 & -0.8325 & 0.0122 \\
2456714.69144 & 0.5675 & 0.0105 & -0.8372 & 0.0120 \\
2456714.69219 & 0.5627 & 0.0107 & -0.8352 & 0.0122 \\
2456714.69293 & 0.5634 & 0.0108 & -0.8452 & 0.0124 \\
2456714.69367 & 0.5678 & 0.0107 & -0.8338 & 0.0122 \\
2456714.69441 & 0.5643 & 0.0110 & -0.8345 & 0.0126 \\
2456715.51976 & 0.5614 & 0.0098 & -0.8257 & 0.0110 \\
2456715.5205 & 0.5699 & 0.0098 & -0.8289 & 0.0110 \\
2456715.52124 & 0.5637 & 0.0097 & -0.8291 & 0.0109 \\
2456715.52199 & 0.5639 & 0.0096 & -0.8246 & 0.0108 \\
2456715.52273 & 0.5665 & 0.0097 & -0.8243 & 0.0108 \\
2457103.52698 & 0.5830 & 0.0107 & -0.8401 & 0.0121 \\
2457103.52773 & 0.5606 & 0.0109 & -0.8414 & 0.0125 \\
2457103.52847 & 0.5753 & 0.0108 & -0.8286 & 0.0122 \\
2457103.52922 & 0.5641 & 0.0109 & -0.8416 & 0.0124 \\
2457103.59318 & 0.5557 & 0.0120 & -0.8272 & 0.0136 \\
2457103.59393 & 0.5453 & 0.0123 & -0.8166 & 0.0139 \\
2457103.59467 & 0.5644 & 0.0119 & -0.8238 & 0.0134 \\
2457103.59541 & 0.5711 & 0.0119 & -0.8368 & 0.0135 \\
2457103.59616 & 0.5536 & 0.0121 & -0.8154 & 0.0136 \\
2457103.65964 & 0.5555 & 0.0146 & -0.8289 & 0.0167 \\
2457103.66039 & 0.5563 & 0.0141 & -0.8228 & 0.0161 \\
2456447.479 & 0.5514 & 0.0104 & -0.8243 & 0.0119 \\
2456448.44667 & 0.8560 & 0.0202 & -0.8277 & 0.0198 \\
2456449.46486 & 0.7808 & 0.0116 & -0.8315 & 0.0119 \\
2456449.49532 & 0.9432 & 0.0142 & -0.8122 & 0.0132 \\
2456450.47442 & 0.5465 & 0.0118 & -0.8306 & 0.0137 \\
2456712.54005 & 0.5602 & 0.0092 & -0.8188 & 0.0103 \\
2456712.59902 & 0.5722 & 0.0093 & -0.8416 & 0.0104 \\
2456712.7069 & 0.5620 & 0.0096 & -0.8396 & 0.0109 \\
2456713.64396 & 0.5776 & 0.0097 & -0.8392 & 0.0109 \\
2456714.52232 & 0.5460 & 0.0106 & -0.8308 & 0.0121 \\
2457103.52624 & 0.5703 & 0.0109 & -0.8294 & 0.0123 \\
2457103.66113 & 0.5620 & 0.0144 & -0.8161 & 0.0164 \\
2457104.57835 & 0.6420 & 0.0174 & -0.8238 & 0.0192 \\
2457103.66188 & 0.5566 & 0.0154 & -0.8136 & 0.0176 \\
2457103.66263 & 0.5532 & 0.0153 & -0.8113 & 0.0174 \\
2457104.52351 & 0.9120 & 0.0127 & -0.8466 & 0.0122 \\
2457104.52426 & 0.8921 & 0.0127 & -0.8498 & 0.0124 \\
2457104.52501 & 0.8979 & 0.0127 & -0.8518 & 0.0124 \\
2457104.52576 & 0.8957 & 0.0134 & -0.8521 & 0.0131 \\
2457104.52652 & 0.8841 & 0.0138 & -0.8395 & 0.0135 \\
2457104.57535 & 0.6505 & 0.0121 & -0.8333 & 0.0132 \\
2457104.5761 & 0.6353 & 0.0138 & -0.8294 & 0.0152 \\
2457104.57685 & 0.6441 & 0.0125 & -0.8341 & 0.0137 \\
2457104.5776 & 0.6343 & 0.0146 & -0.8321 & 0.0161 \\
2457104.5791 & 0.6331 & 0.0211 & -0.8160 & 0.0233 \\
2457104.57986 & 0.6110 & 0.0234 & -0.8288 & 0.0265 \\
2457104.5806 & 0.6277 & 0.0145 & -0.8332 & 0.0161 \\
2457104.58134 & 0.6271 & 0.0138 & -0.8264 & 0.0152 \\
2457104.58209 & 0.6227 & 0.0133 & -0.8199 & 0.0147 \\
2457104.58283 & 0.6148 & 0.0120 & -0.8311 & 0.0133 \\
2457104.58358 & 0.6163 & 0.0124 & -0.8305 & 0.0138 \\
2457104.58432 & 0.6084 & 0.0121 & -0.8286 & 0.0134 \\
2457104.58508 & 0.6205 & 0.0115 & -0.8367 & 0.0128 \\
2457104.58583 & 0.6170 & 0.0113 & -0.8287 & 0.0125 \\
2457104.65942 & 0.5369 & 0.0339 & -0.7894 & 0.0394 \\
2457104.66017 & 0.6066 & 0.0440 & -0.9166 & 0.0538 \\
2457104.66092 & 0.5450 & 0.0357 & -0.8370 & 0.0427 \\
2457104.66166 & 0.5512 & 0.0355 & -0.8501 & 0.0426 \\
2457104.6624 & 0.5270 & 0.0378 & -0.8074 & 0.0448 \\

\hline                                   
\end{longtable}

}

\longtab{2}{
\begin{longtable}{llrrllllll}
 \caption{\label{Tab:lines} Line intensities in the nebula of M3-2}\\
 \hline\hline
 $ \lambda_{\rm Obs}$ & $ \lambda_{\rm True}$ & $F \left( \lambda \right) $ & $I \left( \lambda \right) $ & Ident. & Multiplet & Lower term & Upper term & g$_1$ & g$_2$ \\
 \hline
\endfirsthead
\caption{continued.}\\
\hline\hline
 \hline
 $ \lambda_{\rm Obs}$ & $ \lambda_{\rm True}$ & $F \left( \lambda \right) $ & $I \left( \lambda \right) $ & Ident.  & Multiplet & Lower term & Upper term & g$_1$ & g$_2$ \\
\endhead
\hline
\endfoot
\\
  3630.00 &   3630.00 &    0.171 $\pm$   0.116&   0.279 $^{  +0.117}_{  -0.114}$ & Balmer cont.                                                                       \\
  3700.00 &   3700.00 &    0.264 $\pm$   0.113&   0.252 $^{  +0.114}_{  -0.113}$ & Paschen cont.                                                                      \\
  3724.56 &   3721.63 &   10.042 $\pm$   1.142&  10.000 $\pm$   0.100 & [S~{\sc iii}]    & F2         & 3p2 3P     & 3p2 1S     &          3 &        1    \\
        ... &   3721.94 &  ...                  & ...              & H~{\sc i}        & H14        & 2p+ 2P...    & 14d+ 2D    &          8 &        ...    \\
  3728.97 &   3726.03 &   64.393 $\pm$   1.282&  63.800 $\pm$   1.300 & [O~{\sc ii}]     & F1         & 2p3 4S...    & 2p3 2D...    &          4 &        4    \\
  3731.76 &   3728.82 &   43.271 $\pm$   1.312&  41.900 $\pm$   1.300 & [O~{\sc ii}]     & F1         & 2p3 4S...    & 2p3 2D...    &          4 &        6    \\
  3838.41 &   3835.39 &    7.045 $\pm$   0.921&   6.452 $^{  +0.928}_{  -0.926}$ & H~{\sc i}        & H9         & 2p+ 2P...    & 9d+ 2D     &          8 &        ...    \\
  3871.80 &   3868.75 &  104.646 $\pm$   3.069& 104.000 $\pm$   3.000 & [Ne~{\sc iii}]   & F1         & 2p4 3P     & 2p4 1D     &          5 &        5    \\
  3892.11 &   3889.05 &   21.879 $\pm$   3.006&  20.100 $\pm$   3.000 & H~{\sc i}        & H8         & 2p+ 2P...    & 8d+ 2D     &          8 &        ...    \\
  3970.59 &   3967.46 &   27.571 $\pm$   0.410&  27.840 $\pm$   0.410 & [Ne~{\sc iii}]   & F1         & 2p4 3P     & 2p4 1D     &          3 &        5    \\
  3973.20 &   3970.07 &   16.009 $\pm$   0.413&  16.306 $\pm$   0.411 & H~{\sc i}        & H7         & 2p+ 2P...    & 7d+ 2D     &          8 &       98    \\
  4071.14 &   4067.94 &    6.540 $\pm$   0.792&   6.540 $\pm$   0.010 & C~{\sc iii}      & V16        & 4f 3F...     & 5g 3G      &          5 &        7    \\
        ... &   4068.60 &  ...                  & ...              & [S~{\sc ii}]     & F1         & 2p3 4S...    & 2p3 2P...    &          4 &        4    \\
        ... &   4068.92 &  ...                  & ...              & C~{\sc iii}      & V16        & 4f 3F...     & 5g 3G      &          7 &        7    \\
        ... &   4069.62 &  ...                  & ...              & O~{\sc ii}       & V10        & 3p 4D...     & 3d 4F      &          2 &        4    \\
        ... &   4069.89 &  ...                  & ...              & O~{\sc ii}       & V10        & 3p 4D...     & 3d 4F      &          4 &        6    \\
        ... &   4070.26 &  ...                  & ...              & C~{\sc iii}      & V16        & 4f 3F...     & 5g 3G      &          9 &       11    \\
        ... &   4071.23 &  ...                  & ...              & O~{\sc ii}       & V48a       & 3d 4F      & 4f G5...     &          8 &       10    \\
        ... &   4072.16 &  ...                  & ...              & O~{\sc ii}       & V10        & 3p 4D...     & 3d 4F      &          6 &        8    \\
  4104.97 &   4101.74 &   20.361 $\pm$   1.412&  17.500 $^{  +1.400}_{  -1.500}$ & H~{\sc i}        & H6         & 2p+ 2P...    & 6d+ 2D     &          8 &       72    \\
  4343.89 &   4340.47 &   38.466 $\pm$   0.638&  38.500 $\pm$   0.100 & H~{\sc i}        & H5         & 2p+ 2P...    & 5d+ 2D     &          8 &       50    \\
        ... &   4342.00 &  ...                  & ...             & O~{\sc ii}       & V77        & 3d 2F      & 4f 2[5]...   &          8 &       10    \\
  4366.65 &   4363.21 &    9.935 $\pm$   0.728&   9.233 $^{  +0.708}_{  -0.767}$ & [O~{\sc iii}]    & F2         & 2p2 1D     & 2p2 1S     &          5 &        1    \\
  4475.02 &   4471.50 &    5.461 $\pm$   0.392&   5.711 $^{  +0.379}_{  -0.406}$ & He~{\sc i}       & V14        & 2p 3P...     & 4d 3D      &          9 &       15    \\
  4610.66 &   4607.03 &    0.861 $\pm$   0.273&   0.861 $\pm$   0.001 & [Fe~{\sc iii}]   & F3         & 3d6 5D     & 3d6 3F2    &          9 &        7    \\
        ... &   4607.16 &  ...                  & ...              & N~{\sc ii}       & V5         & 3s 3P...     & 3p 3P      &          1 &        3    \\
  4651.08 &   4647.42 &    1.631 $\pm$   0.444&   1.630 $\pm$   0.010 & C~{\sc iii}      & V1         & 3s 3S      & 3p 3P...     &          3 &        5    \\
        ... &   4649.13 &  ...                  & ...              & O~{\sc ii}       & V1         & 3s 4P      & 3p 4D...     &          6 &        8    \\
        ... &   4650.25 &  ...                  & ...              & C~{\sc iii}      & V1         & 3s 3S      & 3p 3P...     &          3 &        3    \\
        ... &   4650.84 &  ...                  & ...              & O~{\sc ii}       & V1         & 3s 4P      & 3p 4D...     &          2 &        2    \\
        ... &   4651.47 &  ...                  & ...             & C~{\sc iii}      & V1         & 3s 3S      & 3p 3P...     &          3 &        1    \\
  4689.37 &   4685.68 &   52.111 $\pm$   1.792&  53.200 $\pm$   1.800 & He~{\sc ii}      & 3.4        & 3d+ 2D     & 4f+ 2F...    &         18 &       32    \\
  4864.72 &   4861.33 &   98.667 $\pm$   1.517& 100.000 $\pm$   1.500 & H~{\sc i}        & H4         & 2p+ 2P...    & 4d+ 2D     &          8 &       32    \\
  4925.37 &   4921.93 &    2.366 $\pm$   0.235&   2.370 $\pm$   0.010 & He~{\sc i}       & V48        & 2p 1P...     & 4d 1D      &          3 &        5    \\
        ... &   4923.93 &  ...                  & ...              & [Fe~{\sc ii}]    &            & 4s2 4P     & 4p 6P...     &          6 &        4    \\
        ... &   4924.53 &  ...                  & ...              & O~{\sc ii}       & V28        & 3p 4S...     & 3d 4P      &          4 &        6    \\
  4962.37 &   4958.91 &  134.176 $\pm$   2.606& 132.000 $\pm$   3.000 & [O~{\sc iii}]    & F1         & 2p2 3P     & 2p2 1D     &          3 &        5    \\
  5010.34 &   5006.84 &  416.828 $\pm$   4.902& 409.000 $\pm$   5.000 & [O~{\sc iii}]    & F1         & 2p2 3P     & 2p2 1D     &          5 &        5    \\
  5201.53 &   5197.90 &   10.741 $\pm$   0.257&  10.539 $\pm$   0.258                                                                                      \\
  5203.89 &   5200.26 &    4.821 $\pm$   0.262&   4.734 $\pm$   0.263 & [N~{\sc i}]      & F1         & 2p3 4S...    & 2p3 2D...    &          4 &        6    \\
  5415.30 &   5411.52 &    4.750 $\pm$   0.188&   4.996 $\pm$   0.188 & He~{\sc ii}      & 4.7        & 4f+ 2F...    & 7g+ 2G     &         32 &       98    \\
  5521.51 &   5517.66 &    0.800 $\pm$   0.208&   0.866 $^{  +0.206}_{  -0.209}$ & [Cl~{\sc iii}]   & F1         & 2p3 4S...    & 2p3 2D...    &          4 &        6    \\
  5541.47 &   5537.60 &    1.047 $\pm$   0.308&   1.162 $^{  +0.314}_{  -0.313}$ & [Cl~{\sc iii}]   & F1         & 2p3 4S...    & 2p3 2D...    &          4 &        4    \\
  5758.62 &   5754.60 &   30.129 $\pm$   0.672&  28.899 $\pm$   0.671 & [N~{\sc ii}]     & F3         & 2p2 1D     & 2p2 1S     &          5 &        1    \\
  5879.76 &   5875.66 &   30.383 $\pm$   1.347&  28.800 $\pm$   1.300 & He~{\sc i}       & V11        & 2p 3P...     & 3d 3D      &          9 &       15    \\
  6304.15 &   6300.34 &    9.477 $\pm$   0.215&   9.557 $\pm$   0.214 & [O~{\sc i}]      & F1         & 2p4 3P     & 2p4 1D     &          5 &        5    \\
  6314.62 &   6310.80 &    4.566 $\pm$   0.224&   4.570 $\pm$   0.010                                                                                      \\
        ... &   6312.10 &  ...                  & ...              & [S~{\sc iii}]    & F3         & 2p2 1D     & 2p2 1S     &          5 &        1    \\
  6367.63 &   6363.78 &    3.022 $\pm$   0.182&   3.189 $\pm$   0.184 & [O~{\sc i}]      & F1         & 2p4 3P     & 2p4 1D     &          3 &        5    \\
  6552.06 &   6548.10 &  363.246 $\pm$   6.629& 370 $\pm$   7 & [N~{\sc ii}]     & F1         & 2p2 3P     & 2p2 1D     &          3 &        5    \\
  6566.74 &   6562.77 &  291.609 $\pm$  14.177& 298 $\pm$  14 & H~{\sc i}        & H3         & 2p+ 2P...    & 3d+ 2D     &          8 &       18    \\
  6587.48 &   6583.50 & 1145.275 $\pm$  13.560&1130 $\pm$  10 & [N~{\sc ii}]     & F1         & 2p2 3P     & 2p2 1D     &          5 &        5    \\
 \end{longtable}

}
\clearpage
\twocolumn

\end{appendix} 

\begin{thebibliography}{}
\bibitem[Astropy Collaboration (2013)]{2013A&A...558A..33A} Astropy Collaboration, 2013, \aap, 558, 33
\bibitem[Astropy Collaboration (2018)]{2018arXiv180102634T} Astropy Collaboration, 2018, \apj, in press
\bibitem[Balick \& Frank(2002)]{2002ARAA..40..439B} Balick, B. \& Frank, A.\ 2002, ARA\&A, 40, 439 
\bibitem[\protect\citeauthoryear{Boffin}{2015a}]{2015ASPC..493..527B} Boffin H., 2015, 19th European Workshop on White Dwarfs,  527, ASPC..493
\bibitem[\protect\citeauthoryear{Boffin}{2015b}]{2015ebss.book..153B} Boffin H.~M.~J., 2015, in Ecology of Blue Straggler Stars (Boffin, Carraro, Beccari, eds),  Springer, pp. 153--178
\bibitem[Buzzoni et al.(1984)]{efosc2}Buzzoni, B., Delabre, B., Dekker, H., Dodorico, S., Enard, D., Focardi, P., Gustafsson, B., Nees, W., Paureau, J., Reiss, R.\ 1984, ESO Messenger, 38, 9 
\bibitem[Cahn \& Kaler(1971)]{1971ApJS...22..319C} Cahn, J.~H., \& Kaler, J.~B.\ 1971, \apjs, 22, 319 
\bibitem[De Marco(2009)]{2009PASP..121..316D} De Marco, O.\ 2009, PASP, 121, 316 
\bibitem[Foreman-Mackey et al.(2013)]{2013PASP..125..306F} Foreman-Mackey, D., Hogg, D.~W. \& Lang, D., 2013, PASP, 125, 306
\bibitem[Foreman-Mackey(2016]{2016JOSS....1...24F} Foreman-Mackey, D., 2016, JOSS, 1, 24
\bibitem[Gaia Collaboration et al.(2016)]{gaia16} Gaia Collaboration, Prusti, T., de Bruijne, J.~H.~J., et al.\ 2016, \aap, 595, A1
\bibitem[Gaia Collaboration et al.(2018)]{gaia18} Gaia Collaboration, Brown, A.~G.~A., Vallenari, A., et al.\ 2018, \aap, in press -- arXiv:1804.09365
\bibitem[Gray \& Corbally(1994)]{1994AJ....107..742G} Gray, R.~O., \& Corbally, C.~J.\ 1994, \aj, 107, 742 
\bibitem[Hunter(2007)]{matplotlib} Hunter, J.D., 2007, Computing in Science \& Engineering, 9, 90
\bibitem[Jacoby \& Kaler(1989)]{1989AJ.....98.1662} Jacoby, G.H. \& Kaler, J.B.\ 1989, AJ, 98, 1662
\bibitem[\protect\citeauthoryear{Jones \& Boffin}{2017a}]{2017NatAs...1E.117J} Jones D., Boffin H.~M.~J., 2017, Nature Astronomy, 1, 117
\bibitem[Jones \& Boffin(2017b)]{2017MNRAS.466.2034J} Jones, D., \& Boffin, H.~M.~J.\ 2017, \mnras, 466, 2034 
\bibitem[Jones et al.(2015)]{jones15} Jones, D. et al.\ 2015, \aap, 580, 19
\bibitem[Jones et al.(2014)]{jones14} Jones, D. et al.\ 2015, \aap, 562, 89
\bibitem[Jones et al.(2016)]{jones16} Jones, D. et al.\ 2016, \mnras, 455, 3263
\bibitem[Kaler \& Jacoby(1989)]{1989ApJ...345..871} Kaler, J.B. \& Jacoby, G.H.\ 1989, ApJ, 345, 871
\bibitem[Kniazev(2012)]{2012AstL...38..707K} Kniazev, A.~Y.\ 2012, Astronomy Letters, 38, 707 
\bibitem[Maciel(1984)]{1984AAS...55..253M} Maciel, W.~J.\ 1984, \aaps, 55, 253 
\bibitem[Minkowski(1948)]{Minkowski1948} Minkowski, R. 1948, PASP, 60, 386
\bibitem[Miszalski et al.(2009)]{Miszalski2009} Miszalski, B. et al. 2009, A\&A, 505, 249
\bibitem[Miszalski et al.(2012)]{A70} Miszalski, B., Boffin, H.~M.~J., Frew, D.~J., et al.\ 2012, \mnras, 419, 39 
\bibitem[Miszalski et al.(2013)]{2013MNRAS.432.3186M} Miszalski, B., Miko{\l}ajewska, J., \& Udalski, A.\ 2013, \mnras, 432, 3186 
\bibitem[Miszalski et al.(2013)]{Hen2-39} Miszalski, B., Boffin, H.~M.~J., Jones, D., et al.\ 2013, \mnras, 436, 3068 
\bibitem[Miszalski et al.(2018)]{MyCn18} Miszalski, B., Manick, R., Miko{\l}ajewska, J., Van Winckel, H., \& I{\l}kiewicz, K.\ 2018, \pasa, 35, e027
\bibitem[Pickles(1998)]{1998PASP..110..863P} Pickles, A.~J., 1998, PASP, 110, 863
\bibitem[Phillips(2004)]{2004MNRAS.353..589P} Phillips, J.~P.\ 2004, \mnras, 353, 589
\bibitem[Price-Whelan \& Foreman-Mackey (2017)]{2017JOSS....2..357P} Price-Whelan, A. \& Foreman-Mackey, D., 2017, JOSS, 2, 375
\bibitem[Pr{\v s}a et al.(2016)]{2016ApJS..227...29P} {Pr{\v s}a}, A., {Conroy}, K.~E., {Horvat}, M., {Pablo}, H., 
	{Kochoska}, A., {Bloemen}, S., {Giammarco}, J., {Hambleton}, K.~M. \& {Degroote}, P., 2016, \apjs, 227, 29
\bibitem[Shaw \& Kaler(1989)]{1989ApJS...69..495S} Shaw, R.A. \& Kaler, J.B.\ 1989, ApJS, 69, 495
\bibitem[Shortridge etal.(2004)]{STARLINK} Shortridge K., Meyerdierks H., Currie M. J., et al., 2004,
Starlink User Note 86.21. Rutherford Appleton Laboratory
\bibitem[Tylenda et al.(1991)]{1991AAS...89...77} Tylenda, R., Acker, A., Raytchev, B., Stenholm, B., \& Gleizes, F. \ 1991, A\&AS, 89, 77
\bibitem[Van Der Walt, Colbert \& Varoquaux(2011)]{numpy} Van Der Walt, S., Colbert, S.~C. \& Varoquaux, G., 2011, Computing in Science \& Engineering, 13, 22
\bibitem[Wesson et al.(2005)]{2005MNRAS.362..424W} Wesson, R., Liu, X.-W., \& Barlow, M.~J.\ 2005, \mnras, 362, 424 
\bibitem[Wesson, Stock \& Scicluna(2012)]{Neat} Wesson, R., Stock, D.~J. \& Scicluna, P. 2012, MNRAS, 422, 3516
\bibitem[Wesson(2016)]{Alfa} Wesson, R. 2016, MNRAS, 456, 3774
 \end{thebibliography}
\end{document}